\documentclass[english,aps,pra,amsmath,twocolumn,preprintnumber,superscriptaddress,showpacs,notitlepage,nofootinbib]{revtex4-1}
\usepackage[T1]{fontenc}
\usepackage[latin9]{inputenc}
\makeatletter

\usepackage{mathptmx,newtxtext,newtxmath,xspace}
\usepackage{amsbsy,bm,bbold}
\usepackage{graphicx,color,xcolor,epsfig,rotate}
\usepackage{fancyhdr}
\usepackage{gensymb}
\usepackage[colorlinks=true, 
            linkcolor=blue, 
            urlcolor=blue,
            citecolor=blue]{hyperref}
\usepackage{soul}
\newcommand{\iu}{{i\mkern1mu}}

\makeatother

\usepackage{babel}
\begin{document}
\title{Resistivity minimum in diluted metallic magnets}
\author{Zhentao~Wang}
\affiliation{Department of Physics and Astronomy, The University of Tennessee,
Knoxville, Tennessee 37996, USA}
\author{Cristian~D.~Batista}
\affiliation{Department of Physics and Astronomy, The University of Tennessee,
Knoxville, Tennessee 37996, USA}
\affiliation{Quantum Condensed Matter Division and Shull-Wollan Center, Oak Ridge
National Laboratory, Oak Ridge, Tennessee 37831, USA}
\date{\today}
\begin{abstract}
Resistivity minima are commonly seen in itinerant magnets and they
are often attributed to the Kondo effect. However, recent experiments
are revealing an increasing number of materials showing resistivity
minima in the absence of indications of Kondo singlet formation. In a previous work [Z.~Wang, K.~Barros, G.-W.~Chern, D.~L.~Maslov, and C.~D.~Batista, \href{https://doi.org/10.1103/PhysRevLett.117.206601}{Phys. Rev. Lett. {\bf 117}, 206601 (2016)}],
we demonstrated that the Ruderman-Kittel-Kasuya-Yosida (RKKY) interaction
can produce a classical spin liquid state at finite temperature, whose
resistivity increases with decreasing temperature. The classical spin liquid exists over a relatively large temperature window because of the frustrated nature of the RKKY interaction produced
by a 2D electron gas. In this work, we
investigate the robustness of the RKKY-induced resistivity upturn against site dilution, which provides an alternative, and more robust, way of stabilizing the classical spin liquid state down to $T=0$. By using series expansions and stochastic Landau-Lifshitz
dynamics simulation, we show that site dilution competes with thermal
fluctuations and further stabilizes the resistivity upturn, which is accompanied by 
a negative magnetoresistivity due to suppression of the electron-spin scattering. 
\end{abstract}
\pacs{~}
\maketitle

\section{Introduction}

The resistivity minimum of metallic magnets is often associated with the Kondo effect~\citep{Kondo1964}.
Indeed,  the scattering by impurities and defects is the dominant 
dissipative mechanism at sufficiently low  temperature because the phonon population becomes arbitrarily small. The Kondo mechanism consists of 
spin-flip impurity scattering.  
Below the so-called Kondo temperature scale $T_{K}$, the individual magnetic impurities are screened by the conduction electrons, 
and this effect suppresses the correlations between different impurities~\cite{Doniach77}.
However, as it was pointed out by Doniach~\cite{Doniach77}, the Kondo effect can be suppressed if the effective
Ruderman-Kittel-Kasuya-Yosida (RKKY) interaction~\citep{Ruderman,Kasuya,Yosida1957} between different magnetic impurities becomes dominant. This work is motivated by the increasing number of materials that exhibit clear indications of dominant RKKY interaction yet still display a resistivity minimum~\citep{Mallik1998_Gd2PdSi3,Majumdar2000,Majumdar2001,Sampathkumaran2003,
Sengupta2004_RCuAs2,Fritsch2005,Fritsch2006,Nakatsuji06,Jammalamadaka2009_Tb5Si3,Mukherjee2010_Tb4LuSi3,Sakata2011,
Iyer2012_Tb5Si3,Kumar2019_Gd4PtAl,Kumar2019_Tb4PtAl}.


In a previous work~\citep{Wang2016_resistivity}, we demonstrated that
the classical spin liquid state produced by a highly frustrated RKKY interaction enhances the back scattering
process ($\Delta k=2k_{F}$) and generates a resistivity upturn at low temperature.\footnote{A special example of this mechanism is the resistivity minimum in spin ice systems, which were demonstrated in Refs.~\cite{Udagawa12,Chern13}.}
In that work, we focused on the dense limit of one magnetic moment per lattice site (the magnetic moments form a periodic array). In the present work, we consider the diluted case in which the concentration $n_s$ of magnetic moments is smaller than one ($n_s<1$) and the moments are randomly distributed scattering centers.
This scenario applies to multiple materials including SrTiO$_{3}$ based thin films/heterostructures~\citep{Das2014,Han2016,Iglesias_thesis1990}, 
dilute magnetic semiconductors (Ga$_{1-x}$Mn$_{x}$)As~\citep{Matsukura1998,Jungwirth2006},
manganites~\citep{Salamon2001},  and intermetallic compounds~\citep{Mallik1998_Gd2PdSi3,Fritsch2006,Nakatsuji06,Mukherjee2010_Tb4LuSi3,Sakata2011}.
It is then natural to ask if the RKKY interaction can produce a  resistivity upturn when the concentration of magnetic impurities is low.

To address this question, we single out the RKKY mechanism by assuming that the local moments are classical
(the Kondo effect is explicitly excluded). This limit is relevant for compounds with large magnetic moments
or materials in which the ``Kondo exchange'' $J$ between the spin of the conduction electrons and the magnetic impurities is 
ferromagnetic (FM). The assumption of an effective RKKY interaction between the magnetic impurities implies that 
we are working in the  weak-coupling limit $J\eta(\epsilon_{F})\ll1$, where $\eta(\epsilon_{F})$ is the density of states at the Fermi level. 
Two natural consequences of the weak-coupling limit are in order.
First, higher-order spin interactions mediated by the electrons can
be neglected (for example, four-spin interactions beyond the RKKY
level~\cite{Batista16}). Second, the mean free path $l\propto W/J^{2}$ is much larger than the Fermi wavelength $k_F l \gg 1$ ($W$ is the bandwidth).
In three dimensions (3D), this condition leads to metallic behavior of the electrons. In two dimensions (2D), the system exhibits metallic behavior down to very low temperatures below which
Anderson localization becomes relevant  (the localization length depends exponentially
on $l$~\citep{Lee1985_rmp}). Furthermore,
the weak localization is suppressed in the presence of magnetic impurities~\citep{Lee1985_rmp}.
Finally, we do not include electron-electron or electron-phonon
interactions, which are responsible for the positive slope  of the resistivity at high enough 
temperatures (the resistivity minimum arises from the combination of this effect with 
the low-temperature resistivity upturn caused by the scattering with magnetic impurities).

\begin{figure}
\centering
\includegraphics[width=1\columnwidth]{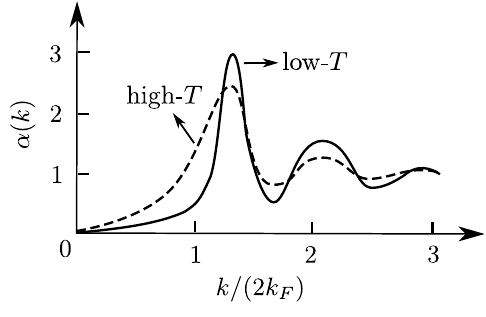}
\caption{Schematic plot of the ionic structure factor $\alpha(k)$ of the liquid metals at different temperatures~\cite{Gingrich61,Ziman61}. \label{fig:Ziman}}
\end{figure}

Another natural question is: how can we distinguish between a resistivity minimum induced by the Kondo effect and the one induced by the RKKY interaction? The results presented in this paper show that the two alternative scenarios can be tested by measuring the temperature dependence of the spin structure factor. When the resistivity minimum is induced by the RKKY interaction, the low-temperature resistivity upturn should be accompanied by an upturn of the magnetic structure factor at wave vectors $k \lesssim 2k_F$. This situation is similar to the electric transport in liquid metals, where the scattering centers are the ionic displacements instead of magnetic impurities~\cite{Ziman61}. Within the Born approximation, the resistance of the liquid metal due to electron-ion scattering is determined by the Fourier transform of the pair distribution function or ionic structure factor $\alpha(k)$. The temperature variation of the resistivity follows from the temperature dependence of $\alpha(k)$, which can be measured with neutron diffraction~\cite{Gingrich61,Ziman61}. For most liquid alkali and noble metals, the resistivity of the liquid is much lower than that of the gas, because the ionic structure factor of the liquid $\alpha_{\rm liq}(k)$ is lower than  that of the gas $\alpha_{\rm gas}(k)={\bar \alpha}$ (${\bar \alpha}$ is the average value of $\alpha(k)$ over the momentum space) for $k \leq 2k_F$ (see Fig.~\ref{fig:Ziman}). This is so because the ionic density is typically larger than the density of conduction electrons, implying that $\alpha_{\rm liq}(k)$ is peaked at a wave vector $K > 2k_F$ (the free electron Fermi surface does not touch the boundary of the Brillouin zone). In other words, the transition from the gas to the liquid is accompanied by a spectral weight transfer from $k \leq  2 k_F$ to $k \simeq K$ that reduces the back scattering by a large amount.

The crucial difference between the above-described  gas to liquid metal crossover and the scattering of electrons by magnetic impurities near the paramagnetic (``spin gas'') to classical spin liquid crossover is that {\it the dominant magnetic correlations of the spin liquid state  are dictated by the electronic density}. In other words, for relatively small Fermi surfaces, the RKKY interaction between magnetic impurities enhances the magnetic structure factor of the classical spin liquid state at $k \lesssim 2k_F$ relative to its value in the high-temperature paramagnetic state: $\mathcal{S}_{\rm liquid}(k\lesssim 2k_F)>\mathcal{S}_{\rm gas}$ (note that $\mathcal{S}_{\rm gas}$ is independent of $k$). This enhancement, that can also be verified with a neutron or x-ray scattering experiment, increases the electronic back scattering and results in a {\it higher resistivity of the classical spin liquid state relative to the high-temperature spin gas}. In contrast, the resistivity upturn produced by the Kondo effect should not be accompanied by a similar upturn of $\mathcal{S}(k\lesssim 2k_F)$.

The rest of this paper is organized as follows. In Sec.~\ref{sec:Model},
we introduce the model and lay out the general formalism of the calculation.
In  Sec.~\ref{sec:dense}, we  discuss the dense limit of two dimensional metals with a small Fermi surface.
Section~\ref{sec:general_filling} is devoted to the general case of  arbitrary concentrations of magnetic impurities, while
Sec.~\ref{sec:dilute} focuses on the dilute limit. In Sec.~\ref{sec:magneto},
we discuss the effect of an applied magnetic field. In Sec.~\ref{sec:ThreeD},
we extend our discussion to three dimensional systems.
Conclusions and further discussions are presented in Sec.~\ref{sec:conclusion}. The 
appendices include the formulas for the high-temperature (high-$T$)
series expansion (Appendix~\ref{sec:High-T}), a description of the stochastic Landau-Lifshitz
(SLL) dynamics (Appendix~\ref{sec:SLL}), a description of the bond-density wave phase that appears at low-enough 
temperature in the dense limit (Appendix~\ref{sec:bond-density}), and a discussion of the finite
size effects in the SLL simulation (Appendix~\ref{sec:Finite-Size}).

\section{Model\label{sec:Model}}

We consider the classical Kondo lattice model (KLM) with uncorrelated
random spin vacancies on a square lattice
\begin{equation}
\mathcal{H}=\sum_{\bm{k}}\sum_{\sigma}\left(\epsilon_{\bm{k}}-\mu\right)c_{\bm{k}\sigma}^{\dagger}c_{\bm{k}\sigma}+J\sum_{i}\sum_{\alpha\beta}c_{i\alpha}^{\dagger}\bm{\sigma}_{\alpha\beta}c_{i\beta}\cdot\zeta_{i}\bm{S}_{i}.\label{eq:Kondo}
\end{equation}
The operators $c_{i\sigma}^{\dagger}$/$c_{i\sigma}$ create/annihilate an electron with spin $\sigma$
on site $i$, while $c_{\bm{k}\sigma}^{\dagger}$/$c_{\bm{k}\sigma}$
are the corresponding operators in Fourier space. $\epsilon_{\bm{k}}$
is the bare electron dispersion with chemical potential $\mu$. $J$
is the exchange interaction between the local magnetic moments $\bm{S}_{i}$
and the conduction electrons ($\bm{\sigma}$ is the vector of the
Pauli matrices). The classical moments are normalized, $\left|\bm{S}_{i}\right|=1$, because the magnitude of the local magnetic moments can be absorbed in the coupling constant $J$.
The choice of a square lattice is immaterial because the underlying lattice geometry does not
alter the frustrated nature of the RKKY interaction in the long wavelength limit $k_F \ll 1$.

The uncorrelated random integers $\zeta_{i}=\{0,1\}$ denote the absence/presence
of a magnetic impurity on site $i$. For a given spin concentration $n_{s}\in[0,1]$,
the uncorrelated random integers $\zeta_{i}$ are drawn from the distribution
\begin{equation}
P(\zeta_{i})=n_{s}\delta_{\zeta_{i},1}+\left(1-n_{s}\right)\delta_{\zeta_{i},0}.
\end{equation}
Consequently, the disorder average of $\zeta_{i}$ gives $\overline{\zeta_{i}}=n_{s}$.

We will focus on the weak-coupling limit $J\eta(\epsilon_{F})\ll1$
of Eq.~(\ref{eq:Kondo}). In this limit, the conduction electrons can be integrated out  to
obtain the effective spin Hamiltonian known as RKKY model~\citep{Ruderman,Kasuya,Yosida1957}:
\begin{equation}
\begin{split}\mathcal{H}_{\text{RKKY}} & =-J^{2}\sum_{\bm{k}}\tilde{\chi}_{\bm{k}}\bm{S}_{\bm{k}}\cdot\bm{S}_{-\bm{k}}\\
 & =\sum_{i,j}J(\bm{r}_{i}-\bm{r}_{j})\left(\zeta_{i}\bm{S}_{i}\right)\cdot\left(\zeta_{j}\bm{S}_{j}\right),
\end{split}
\label{eq:RKKY}
\end{equation}
with
\begin{align}
\tilde{\chi}_{\bm{k}} & =\chi_{\bm{k}}^{0}-\frac{1}{N}\sum_{\bm{k}}\chi_{\bm{k}}^{0},\\
\chi_{\bm{k}}^{0} & =-\frac{1}{2\pi^{2}}\int d\bm{q}\frac{f(\epsilon_{\bm{q}+\bm{k}})-f(\epsilon_{\bm{q}})}{\epsilon_{\bm{q}+\bm{k}}-\epsilon_{\bm{q}}},\label{eq:chi0}\\
\bm{S}_{\bm{k}} & =\frac{1}{\sqrt{N}}\sum_{i}e^{\iu\bm{k}\cdot\bm{r}_{i}}\zeta_{i}\bm{S}_{i},\label{eq:Fourier_S}\\
J(\bm{r}) & =-\frac{J^{2}}{N}\sum_{\bm{k}}e^{\iu\bm{k}\cdot\bm{r}}\tilde{\chi}_{\bm{k}},
\end{align}
where $N=L^{2}$ is the total number of lattice sites, and $f(\epsilon)$
is the Fermi distribution function. In the weak-coupling limit, the Curie-Weiss temperature $\theta_{\rm CW}$ associated with the RKKY interaction is orders of magnitude smaller than the Fermi temperature. 
Correspondingly, we can safely set $T=0$ in the Fermi distribution
function that appears in Eq.~(\ref{eq:chi0}).

In addition, we will focus on the long wave length limit that is obtained for a sufficiently small  Fermi surface (FS).
For concreteness we will fix the  chemical potential at $\mu=-3.396t$ (unless specified otherwise).
We will consider two types of dispersion {relations}. The first one is the parabolic
dispersion
\begin{equation}
\epsilon_{\bm{k}}=-4t+tk^{2},\label{eq:disp_parabolic}
\end{equation}
that is obtained in the long wavelength limit. The  second case corresponds to the 
dispersion relation obtained for the nearest-neighbor tight-binding model
\begin{equation}
\epsilon_{\bm{k}}=-2t\left(\cos k_{x}+\cos k_{y}\right).\label{eq:disp_TB}
\end{equation}

For the tight-binding model, our choice of the chemical potential ($\mu=-3.396t$)
leads to an electron filling fraction of $0.05$  and a small FS (Fermi wave vector
$k_{F}\approx0.8$). The parabolic dispersion is  a very good approximation in this case and its simplicity 
becomes useful for understanding different aspects of the problems that we will consider in this paper.
However, as we will see below, the lattice effects included in the tight-binding dispersion  relation \eqref{eq:disp_TB} change the qualitative behavior
of  the bare electronic susceptibility $\chi_{\bm{k}}^{0}$. The parabolic dispersion leads to a well-known flat maximum of $\chi_{\bm{k}}^{0}$ for $0 \leq k \leq 2 k_F$. This degeneracy is removed by the quartic and higher-order corrections that appear in the Taylor expansion of Eq.~\eqref{eq:disp_TB}. This is the main reason for considering both dispersion relations  in the rest of the manuscript. As we explain in Sec.~\ref{sec:conclusion}, there are multiple physical mechanisms that can produce an effective magnetic interaction between the local moments that is qualitatively the same as the one obtained for the tight-binding model.  
Consequently, the results that we will present for the RKKY interaction derived from Eq.~\eqref{eq:disp_TB} are representative of more general situation in which the magnetic susceptibility at $k=2 k_F$ is slightly higher than the ferromagnetic susceptibility.

The bare electronic susceptibility for the parabolic dispersion \eqref{eq:disp_parabolic} is~\citep{Giuliani_book}
\begin{equation}
\chi_{\bm{k}}^{0}=\frac{1}{2\pi t}\left[\Theta(1-x)+\Theta(x-1)\left(1-\sqrt{1-x^{-2}}\right)\right],\label{eq:chi_parabolic}
\end{equation}
where $x\equiv k/(2k_{F})$. In the thermodynamic limit $N\rightarrow\infty$,
the corresponding real-space RKKY interaction is
\begin{align}
J(\bm{r}) & =\frac{J^{2}}{t}\left(-\frac{k_{F}^{2}}{4\pi^{3/2}}\right)G_{1,3}^{2,0}\left(k_{F}^{2}r^{2}\left|\begin{array}{ccc}
\frac{1}{2}\\
0, & 0, & -1
\end{array}\right.\right)\label{eq:rkky_Meijer}\\
 & \stackrel{k_{F}r\gg1}{\approx}-\frac{J^{2}}{t}\frac{\sin\left(2k_{F}r\right)}{4\pi^{2}r^{2}},
\end{align}
where $G_{pq}^{mn}\left(z\left|\begin{array}{ccc}
a_{1}, & \cdots, & a_{p}\\
b_{1}, & \cdots, & b_{q}
\end{array}\right.\right)$ is the Meijer G function.

The bare electronic susceptibility of the tight-binding model (\ref{eq:disp_TB})
is also known analytically along the high-symmetry directions~\citep{Benard1993}.
Given that $\chi_{\bm{k}}^{0}$ must be evaluated for arbitrary values of $\bm{k}$, 
we will solve  Eq.~(\ref{eq:chi0}) by applying numerical integration methods~\citep{Hahn2005_cuba}.

\begin{figure}
\centering
\includegraphics[width=1\columnwidth]{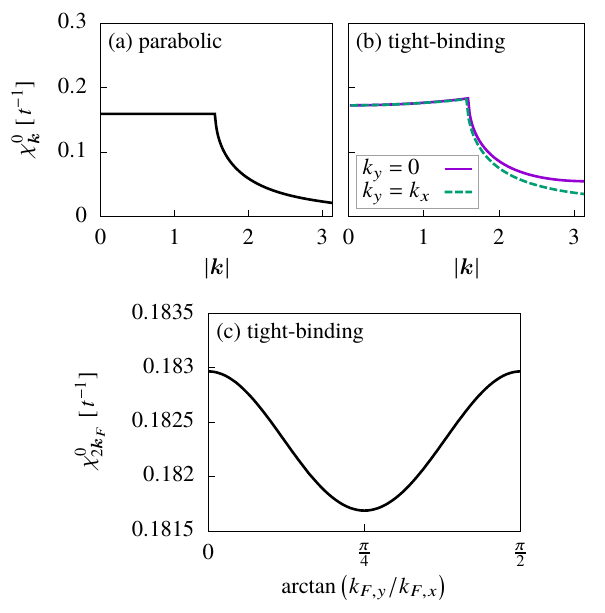}
\caption{Bare electronic susceptibilities for a 2D electron gas. (a) For parabolic
dispersion along radial direction. (b) For tight-binding dispersion
along radial direction. (c) Angular dependence for tight-binding dispersion  at $k=2k_{F}$. \label{fig:chi0}}
\end{figure}

Figure~\ref{fig:chi0} shows the bare electronic susceptibility $\chi_{\bm{k}}^{0}$
for a small FS in 2D. 
$\chi_{\bm{k}}^{0}$ is perfectly flat below $2k_{F}$ for the parabolic dispersion \eqref{eq:disp_parabolic}; 
while a minor upturn from $k=0$ to 
 $k=2k_{F}$ appears for the tight-binding dispersion \eqref{eq:disp_TB}, along with a small angular modulation.
Both effects are caused by quartic, $\mathcal{O}(k^{4})$, corrections to the parabolic dispersion.
Interestingly enough, the small upturn at $2k_{F}$ is enough to  stabilize 
various magnetic phases including spiral, conical and skyrmion crystal orderings at low enough  temperatures~\citep{Wang2020_RKKYskx}.

Within the Born approximation, the inverse relaxation time for elastic
scattering is given by
\begin{equation}
\frac{1}{\tau_{\bm{k}_{F}}}=\frac{4\pi J^{2}}{N}\sum_{\bm{k}}\delta(\mu-\epsilon_{\bm{k}})\overline{\mathcal{S}}(\bm{k}-k_{F})\left(1-\cos\theta_{\bm{k}_{F},\bm{k}}\right),\label{eq:relaxation_time}
\end{equation}
where $\overline{\mathcal{S}}(\bm{k})$ is the disorder averaged static
spin structure factor
\begin{equation}
\overline{\mathcal{S}}(\bm{k})=\overline{\langle\bm{S}_{\bm{k}}\cdot\bm{S}_{-\bm{k}}\rangle},
\end{equation}
and $\langle\cdots\rangle$ denotes thermal average.

The electrical resistivity is given by
\begin{equation}
\rho=\rho_\text{RKKY}(T)=\frac{4}{\pi}\rho_{0}\int_{0}^{1} d x\frac{x^{2}}{\sqrt{1-x^{2}}}\overline{\mathcal{S}}(2k_{F}x),\label{eq:rho_born}
\end{equation}
where $\rho_{0}=2\pi J^{2}/(tek_{F})^{2}$. It is clear then that the transport  cross section due to
electron-spin scattering is determined by the behavior of  the static spin structure factor $\overline{\mathcal{S}}(\bm{k})$
for $k \leq 2 k_F$. Especially, the back scattering process
with $k=2k_{F}$ has the largest weight in Eq.~(\ref{eq:rho_born}).

As we discussed in Ref.~\cite{Wang2016_resistivity}, Eq.~\eqref{eq:rho_born} results from  expanding the  $T$ matrix to the lowest order, while the Kondo effect comes from the spin-flip second-order process (for {\it quantum} spins):
\begin{equation}
\rho \approx \rho_\text{RKKY}(T) \cdot \left[ 1- 8J \eta(\epsilon_F) \ln \left( \frac{T}{D} \right)\right],
\end{equation}
where $D$ is the bandwidth. The usual expression given by Kondo~\cite{Kondo1964} is recovered in the single-impurity limit because  $\rho_\text{RKKY}(T)$ becomes temperature independent (the magnetic structure factor is momentum independent for a single impurity). In contrast, as it is shown in Ref.~\cite{Wang2016_resistivity} and in this paper, $\rho_\text{RKKY}(T)$ can also produce a resistivity upturn for a finite magnetic impurity concentration. To single out this effect, in this paper, we only consider {\it classical} spins so that the Kondo effect is explicitly excluded.

In the following sections, we will evaluate Eq.~\eqref{eq:rho_born} by using 
the average spin structure factor $\overline{\mathcal{S}}(\bm{k})$
obtained from several approximate analytical methods which are appropriate
for different physical limits. We also provide unbiased numerical
results that are obtained by integrating the SLL equation~\citep{LandauLifshitz1992,Gilbert2004}
of the RKKY model.

\section{Dense limit $n_{s}=1$\label{sec:dense}}

The dense limit ($n_{s}=1$) has been discussed briefly in Ref.~\citep{Wang2016_resistivity},
which observed that highly frustrated RKKY interaction leads to a
resistivity upturn upon lowering temperature. Here, we further explore
how the competition between entropy and energy determines the fate
of the upturn at even lower temperatures.

As it is explicitly shown in Appendix~\ref{sec:High-T}, the disorder averaged static spin structure factor
for arbitrary spin concentration can be analytically expanded in powers of the dimensionless parameter 
 $\frac{K}{t}\equiv\frac{2\beta}{3}\frac{J^{2}}{t}n_{s}$ (high-temperature expansion):
\begin{align}
\frac{\overline{\mathcal{S}}(\bm{k})}{n_{s}} & \approx 1+K\tilde{\chi}_{\bm{k}}+K^{2}\left(\tilde{\chi}_{\bm{k}}^{2}-\frac{1}{N}\sum_{\bm{q}}\tilde{\chi}_{\bm{q}}^{2}\right)\nonumber \\
 & \quad+K^{3}\Bigg[\tilde{\chi}_{\bm{k}}^{3}-\frac{1}{N}\sum_{\bm{q}}\tilde{\chi}_{\bm{q}}^{3}-\frac{2}{N}\tilde{\chi}_{\bm{k}}\sum_{\bm{q}}\tilde{\chi}_{\bm{q}}^{2}\nonumber \\
 & \quad\qquad+\left(1-\frac{3}{5n_{s}^{2}}\right)\frac{1}{N^{2}}\sum_{\bm{q}\bm{q}^{\prime}}\tilde{\chi}_{\bm{q}}\tilde{\chi}_{\bm{q}^{\prime}}\tilde{\chi}_{\bm{k}-\bm{q}-\bm{q}^{\prime}}\Bigg].\label{eq:high-T}
\end{align}
By setting $n_{s}=1$ we recover the known result in the dense limit~\citep{Wang2016_resistivity},
which is the focus of the current section. Note that 
all disorder realizations become the same for $n_{s}=1$, so the ``disorder average''
$\overline{\cdots}$ does not change anything in this limit. 

\begin{figure}
\centering
\includegraphics[width=1\columnwidth]{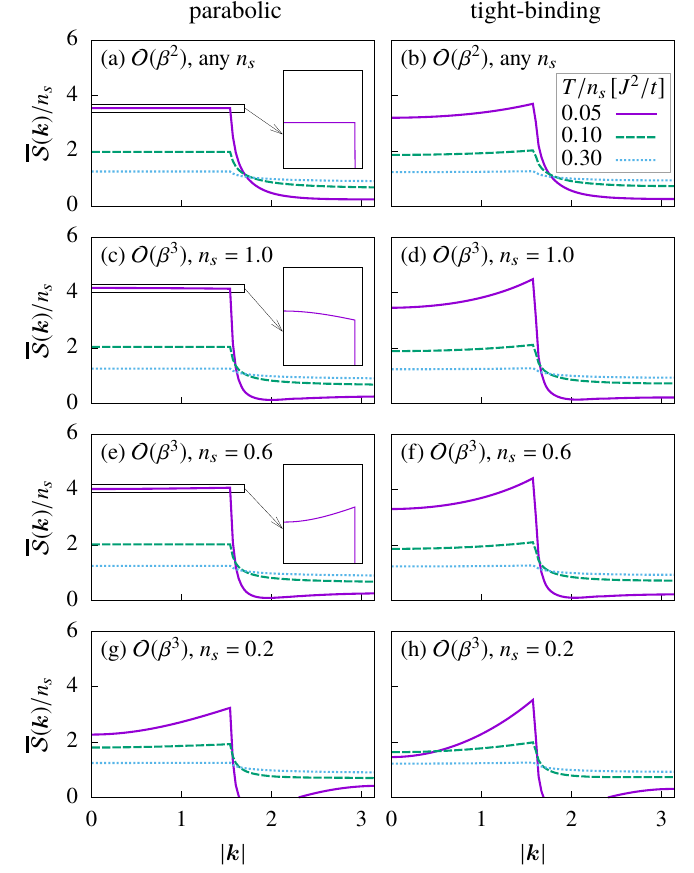}
\caption{Disorder averaged static spin structure factors obtained from high-$T$
expansion (\ref{eq:high-T}), for the parabolic dispersion  \eqref{eq:disp_parabolic} (left column)
and tight-binding dispersion \eqref{eq:disp_TB} (right column) and three different temperatures
$T/n_{s}=\{0.05,~0.1,~0.3\}$ (unit: $J^{2}/t$). [(a) and (b)] include contributions
up to $\mathcal{O}(\beta^{2})$ for any spin concentration; [(c)--(h)]
include contributions up to $\mathcal{O}(\beta^{3})$ for three different
spin concentrations $n_{s}=\{1.0,~0.6,~0.2\}$. The insets show the
zoomed in view of the $T/n_{s}=0.05J^{2}/t$ curves. The angular dependence on $\bm{k}$
is negligible for low electron
filling and high temperature: $\overline{\mathcal{S}}(\bm{k})\approx\overline{\mathcal{S}}(k)$.\label{fig:Sq_highT}}
\end{figure}

\begin{figure}
\centering
\includegraphics[width=1\columnwidth]{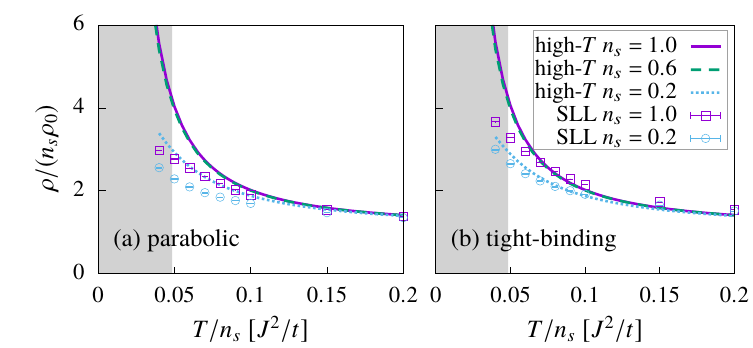}
\caption{Electrical resistivity obtained from high-$T$ expansion up to order
$\mathcal{O}(\beta^{3})$, for (a) parabolic dispersion, and (b) tight-binding
dispersion. We use three different spin concentrations $n_{s}=\{1.0,~0.6,~0.2\}$.
The high-$T$ expansion becomes unreliable in the shaded low-temperature region.
The symbols correspond to the results obtained from SLL simulations of the RKKY model on a square lattice of $128\times 128$ sites.\label{fig:rho_highT}}
\end{figure}

We start by considering the parabolic case, for which $\chi_{\bm{k}}^{0}$ is given
by Eq.~(\ref{eq:chi_parabolic}). Figures~\ref{fig:Sq_highT}(a) and \ref{fig:Sq_highT}(c)
show that $\overline{\mathcal{S}}(\bm{k})$ is enhanced for $k<2k_{F}$
as the temperature is lowered. The increasing weight of $\overline{\mathcal{S}}(\bm{k})$
below $2k_{F}$ significantly enhances the electron-spin scattering
{[}see Eq.~(\ref{eq:rho_born}){]}, and leads to the resistivity upturn
shown in Fig.~\ref{fig:rho_highT}(a). This result is  consistent
with Ref.~\citep{Wang2016_resistivity}.

\begin{figure}
\centering
\includegraphics[width=1\columnwidth]{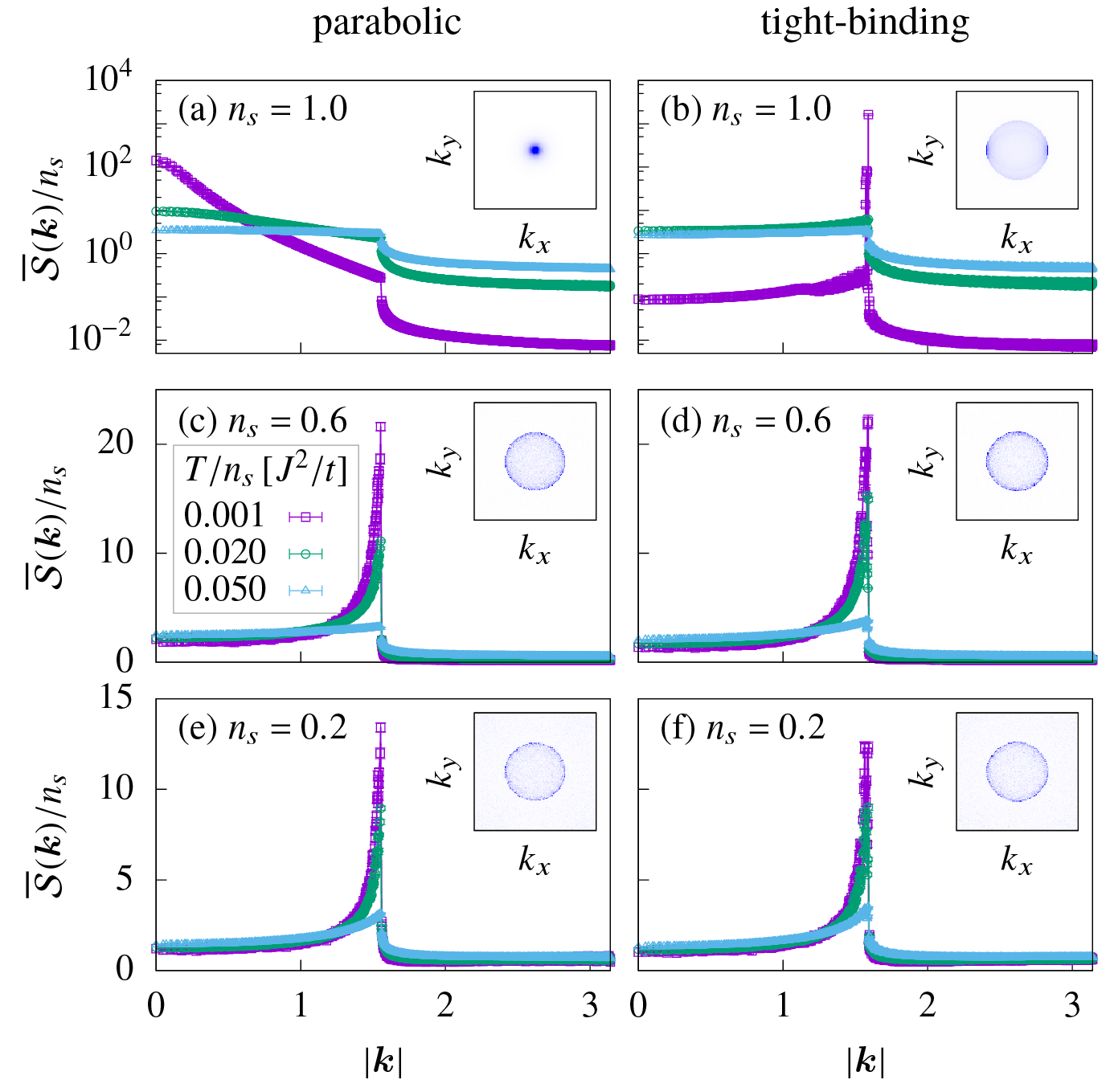}
\caption{Disorder averaged static spin structure factors obtained from SLL
simulation of the RKKY model on $L=128$ square lattice, for the parabolic
dispersion \eqref{eq:disp_parabolic} (left column) and tight-binding dispersion \eqref{eq:disp_TB} (right column).
We use three different spin concentrations $n_{s}=\{1.0,~0.6,~0.2\}$
and three different temperatures $T/n_{s}=\{0.001,0.02,0.05\}$ (unit:
$J^{2}/t$). In the main panels, we average over 64 disorder realizations
to estimate the error bars. The insets show the distribution of $\mathcal{S}(\bm{k})$
in the 1st BZ for one disorder realization at $T/n_{s}=0.001J^{2}/t$
(linear color scales are used in all insets).\label{fig:Sq_SLL}}
\end{figure}

\begin{figure}
\centering
\includegraphics[width=1\columnwidth]{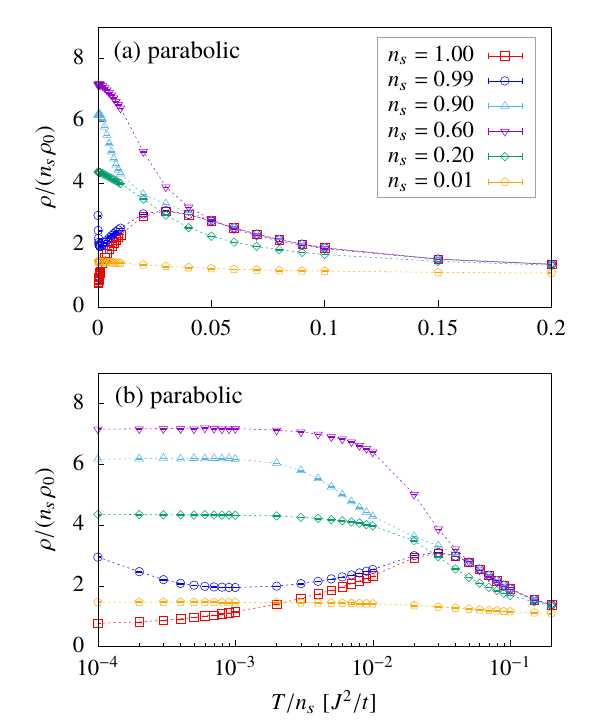}
\caption{Electrical resistivity obtained from SLL simulation of the RKKY model
on a square lattice of $128\times 128$ sites for the susceptibility $\chi^0_{\bm k}$ obtained from the parabolic dispersion \eqref{eq:disp_parabolic}. The error bars are estimated by averaging over 64 realizations of disorder. Both panels
show the same data, with (a) linear scale and (b) linear-logarithmic scale.\label{fig:rho_SLL_parabolic}}

\end{figure}

A closer look at Eq.~\eqref{eq:high-T} reveals that $\overline{\mathcal{S}}(\bm{k})$
is perfectly flat below $2k_{F}$ to order $\mathcal{O}(\beta^{2})$
{[}see Fig.~\ref{fig:Sq_highT}(a){]}. 
This frustration (degeneracy) is eventually lifted by thermal fluctuations,
namely the ``order by disorder'' mechanism~\citep{villain80}. Up
to order $\mathcal{O}(\beta^{3})$, Figure \ref{fig:Sq_highT}(c)
shows that the wave vector $\bm{k}=0$ is entropically favored in comparison 
to any other finite $\bm{k}_{0} \leq 2k_{F}$. The high-$T$ expansion is no longer reliable for very low temperatures
($T/n_{s}<0.05$),  as it is clear from the unphysical negative values of the spin structure factor. However,
our unbiased SLL simulations demonstrate that the weight at $\bm{k}=0$
becomes finally dominant, showing a tendency towards ferromagnetic
ordering {[}see Fig.~\ref{fig:Sq_SLL}(a){]}. According to Eq.~(\ref{eq:rho_born}), it is  clear that the electron-spin
scattering must be suppressed at the lowest temperatures. Indeed, as it is shown in Fig.~\ref{fig:rho_SLL_parabolic} for $n_{s}$=1, the resistivity curve has a low-temperature maximum, i.e., the upturn saturates and it becomes a downturn upon further reducing temperature.

We now turn to the discussion of the tight-binding dispersion. Similar
to the parabolic case, $\chi_{\bm{k}}^{0}$ is also highly frustrated
below $2k_{F}$ {[}see Fig.~\ref{fig:chi0}(b){]}. Upon lowering
temperature, the spectral weight of $\overline{\mathcal{S}}(\bm{k})$
is enhanced  below $2k_{F}$ {[}see Figs.~\ref{fig:Sq_highT}(b) and \ref{fig:Sq_highT}(d){]},
and a resistivity upturn is induced by exactly the same mechanism
{[}see Fig.~\ref{fig:rho_highT}(b){]}. Unlike the parabolic case, $\chi_{\bm{k}}^{0}$ is not perfectly flat
below $2k_{F}$ {[}see Fig.~\ref{fig:chi0}(b){]}. As we already mentioned,
the difference arises  from quartic corrections to the parabolic dispersion \eqref{eq:disp_parabolic}~\citep{Wang2020_RKKYskx}. As shown in
Fig.~\ref{fig:chi0}(b), these corrections induce an upturn along the radial direction, which produces a maximum of 
$\chi_{\bm{k}}^{0}$ at $k=2k_{F}$. In other words, the quartic corrections partially remove the magnetic frustration in  favor of the ring of wave vectors with  $k=2k_{F}$. This different selection mechanism explains the different behaviors of $\overline{\mathcal{S}}(\bm{k})$ for the parabolic and tight-binding cases shown in Figs.~\ref{fig:Sq_highT}(a)--\ref{fig:Sq_highT}(d).
The dominance of the energetic contribution over the entropic one,
i.e., maxima at $2k_{F}$ instead of at $\bm{k}=0$ for $\overline{\mathcal{S}}(\bm{k})$,
is quite clear even for moderate temperatures {[}see Figs.~\ref{fig:Sq_highT}(b) and \ref{fig:Sq_highT}(d){]}.

\begin{figure}
\centering
\includegraphics[width=1\columnwidth]{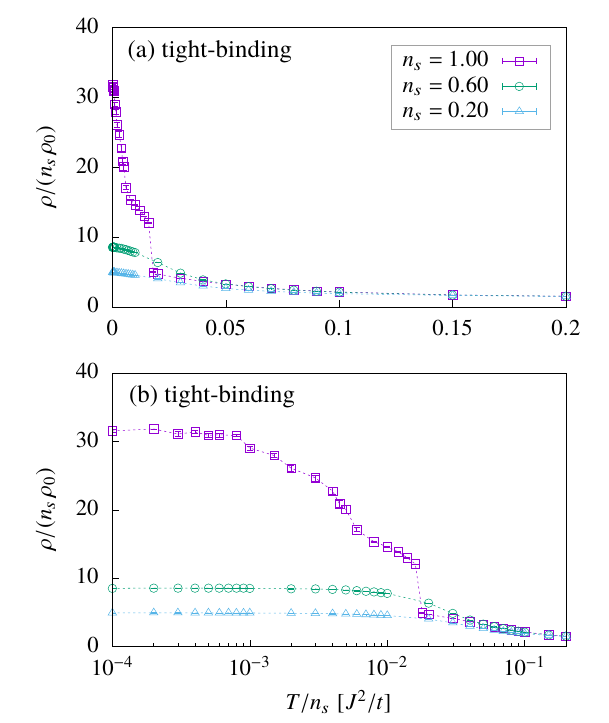}
\caption{Electrical resistivity obtained from SLL simulation of the RKKY model
on square lattice of $128\times 128$ sites  for the susceptibility $\chi^0_{\bm k}$ obtained from the tight-binding dispersion \eqref{eq:disp_TB}. The error bars are estimated by averaging over 64 realizations of disorder .
Both panels show the same data, with (a) linear scale and (b) linear-logarithmic
scale.\label{fig:rho_SLL_square}}
\end{figure}

While the high-$T$ expansion is not reliable at low enough temperatures,
the internal energy contribution should become even more dominant than the entropic contribution upon further reducing $T$.
Indeed, our SLL simulation confirms the development of a sharp  maximum at $k=2k_{F}$ along the radial direction {[}see Fig.~\ref{fig:Sq_SLL}(b){]}. Since $\overline{\mathcal{S}}(2k_F)$
keeps growing upon further lowering
the temperature, the Born approximation
Eq.~(\ref{eq:rho_born}) predicts that the resistivity upturn should persist
for $T \to 0$ (see Fig.~\ref{fig:rho_SLL_square},
$n_{s}=1$ curve), in contrast to the nonmonotonic behavior that was obtained for the parabolic dispersion.
Another interesting consequence of the quartic-correction to the parabolic dispersion is the emergence of a bond-ordered phase at low enough temperatures (see Appendix~\ref{sec:bond-density}) via a first order phase transition. This is the origin of the discontinuous behavior of $\rho(T)$ at $T \simeq 0.02 J^2/t$ for $n_s=1$ that is shown in Fig.~\ref{fig:rho_SLL_square}.

\section{Generic filling $n_{s}\le1$\label{sec:general_filling}}

In this section, we move away from the dense limit to analyze the behavior of the
resistivity for general values of $n_{s}$. We will start with a few general observations that result from the high-$T$ expansion in Eq.~(\ref{eq:high-T}).
First, up to $\mathcal{O}(\beta^{2})$, the temperature dependence
of the $\overline{\mathcal{S}}(\bm{k})$ is the same for any $n_{s}$,
up to an overall rescaling of $\overline{\mathcal{S}}(\bm{k})$ and
 $T$ {[}see Figs.~\ref{fig:Sq_highT}(a) and \ref{fig:Sq_highT}(b){]}. Consequently,
the resistivity upturn should always appear at high enough temperature regardless
of the choice of $n_{s}$. This observation is confirmed by the results shown in Fig.~\ref{fig:rho_highT}.
Second, the overall rescaling factor $n_{s}$ in front of  $\overline{\mathcal{S}}(\bm{k})$
simply means that the cross section is proportional to the number of magnetic impurities  {[}see Eq.~(\ref{eq:rho_born}){]},
implying that  $\rho$ is also rescaled  by $n_{s}$. 
The qualitative behavior of $\rho(T)$ remains the same as the one obtained for the dense limit (see Fig.~\ref{fig:rho_highT}). Third,  the first  nontrivial 
correction of order $\mathcal{O}(\beta^{3})$ changes the behavior of the  resistivity
at low temperatures. 

For the parabolic dispersion, 
the only term of  order $\mathcal{O}(\beta^{3})$ with  momentum dependence for $k<2k_{F}$ has a coefficient
$\left[1-3/\left(5n_{s}^{2}\right)\right]$. This coefficient has 
opposite signs in the high and low spin concentration limits, implying that the dominant spin-spin correlations
become ferromagnetic ($\bm{k}=0$) for $n_{s}=1$ {[}see Fig.~\ref{fig:Sq_highT}(c){]},
and antiferromagnetic with $k=2k_{F}$ for $n_{s}\ll1$ {[}see Figs.~\ref{fig:Sq_highT}(e) and \ref{fig:Sq_highT}(g){]}.
The critical spin concentration for the sign change is $n_{s}=\sqrt{3/5}\approx0.77$.
Note that this critical concentration corresponds to a crossover between two different 
high-temperature behaviors. While our SLL simulation
does not have enough statistics to pin down the critical
concentration for $T \to 0$, we have unambiguously verified that the dominant correlations are ferromagnetic for
$n_{s}\ge0.9$ and antiferromagnetic with $k=2k_F$  for $n_{s}\le0.7$ {[}see also Figs.~\ref{fig:Sq_SLL}(a), \ref{fig:Sq_SLL}(c), and \ref{fig:Sq_SLL}(e){]}.

The different behaviors that are obtained in the low-temperature regime for a parabolic dispersion
and $n_s \lesssim 1$ reflect a characteristic feature of highly frustrated systems: weak effective interactions can tip the balance in one way or another. The dominant ferromagnetic correlations that are obtained in the dense limit arise from a rather fragile order by disorder mechanism induced by thermal fluctuations. In contrast, as we will see in the next section, the dominant $k=2k_F$ correlations for $n_{s}\ll 1$ arise from a much more robust and generic mechanism that has its roots in the oscillatory nature of the RKKY interaction. This phenomenon leads to a more pronounced resistivity upturn  that holds down to very low temperatures (see Figs.~\ref{fig:rho_SLL_parabolic} and \ref{fig:rho_SLL_square}).

In summary, for a parabolic or tight-binding dispersion, the slope of the resistivity is negative, $d\rho /dT <0$, for any $n_{s}$ in the high-temperature regime  because the spectral weight of $\overline{\mathcal{S}}(\bm{k})$
is transferred from the region $k>2k_F$ to the region $k \leq 2k_{F}$ upon decreasing $T$. A further reduction of the temperature leads to the onset of  ferromagnetic correlations for $n_s \simeq 1$ and a parabolic dispersion, while 
antiferromagnetic correlations  ($k=2k_{F}$) become dominant for $n_{s}< 0.7$
(see Figs.~\ref{fig:Sq_highT} and \ref{fig:Sq_SLL}). The natural consequence of these different behaviors is that the resistivity 
has a maximum at low enough temperature for a parabolic dispersion and $n_{s} \simeq 1$, while 
it saturates towards the  $T \to 0$ value for $n_{s} < 0.7$ (see Fig.~\ref{fig:rho_SLL_parabolic}).

The tight-binding dispersion produces a similar behavior, except for the low-temperature regime of the high density limit.
The basic difference is that $\chi^0_{\bm{k}}$ is now maximized at $k=2k_F$, implying that the dominant correlations remain antiferromagnetic  down to $T=0$ even in the $n_{s}=1$ limit. This difference reflects the fragility of the order by disorder mechanism that selects the ferromagnetic correlations for the highly-frustrated RKKY Hamiltonian produced by a  parabolic dispersion. Small perturbations, such as the quartic correction to the parabolic dispersion, can partially release the frustration in favor of ferromagnetic or antiferromagnetic correlations. For instance,  a negative quartic correction, like the one that is obtained for the nearest-neighbor tight-binding dispersion that we are considering, leads to dominant low-temperature antiferromagnetic correlations, while a positive quartic correction leads to dominant ferromagnetic correlations ($\chi^0_{\bm{k}}$ has a global maximum at ${\bm k}={\bm 0}$).


\section{Dilute limit $n_{s}\ll1$\label{sec:dilute}}

We can further understand the dilute limit $n_{s}\ll1$ through a
different perturbative treatment~\citep{Elliott1962,Rushbrooke1964}.
To lowest order in $n_{s}$, we only need to consider two local spins,
located at $\bm{R}_{0}$ and $\bm{R}_{1}$ ($\bm{R}_{0}\neq\bm{R}_{1}$).
The spin-spin correlator can be obtained exactly in this limit:
\begin{equation}
\langle\bm{S}_{\bm{R}_{0}}\cdot\bm{S}_{\bm{R}_{1}}\rangle=\frac{1}{2\beta J(\bm{R}_{0}-\bm{R}_{1})}-\coth\left[2\beta J(\bm{R}_{0}-\bm{R}_{1})\right].
\label{tsc}
\end{equation}

\begin{figure}
\centering
\includegraphics[width=1\columnwidth]{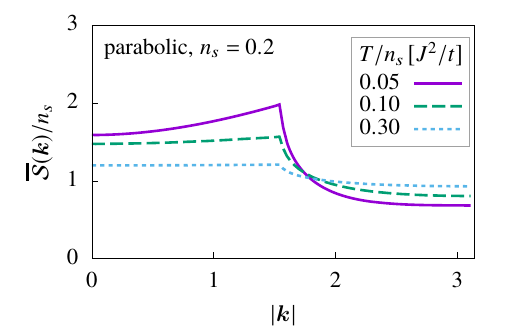}
\caption{Disorder averaged static spin structure factors obtained from low-density
expansion (\ref{eq:Sq_lowNS}) for the parabolic dispersion \eqref{eq:disp_parabolic} with spin
concentration  $n_s=0.2$ and three different temperatures $T/n_{s}=\{0.05,\,0.1,\,0.3\}$
(unit: $J^{2}/t$).\label{fig:Sq_lowNS}}

\centering
\includegraphics[width=1\columnwidth]{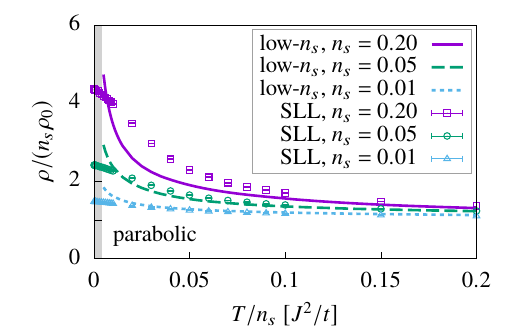}
\caption{Electrical resistivity obtained from low-density expansion (\ref{eq:rho_lowNS})
for the parabolic dispersion \eqref{eq:disp_parabolic} and three different spin concentrations
$n_{s}=\{0.2,\,0.05,\,0.01\}$. The structure factor $\overline{\mathcal{S}}(\bm{k})$ obtained from the expansion is not positive-definite in the shaded low-$T$ region. The symbols show results obtained from SLL simulation, where the parameters are the same as in  Fig.~\ref{fig:rho_SLL_parabolic}.}
\label{fig:rho_lowNS}
\end{figure}

To order $\mathcal{O}(n_{s}^{2})$, the disorder averaged spin structure
factor is
\begin{equation}
\overline{\mathcal{S}}(\bm{k})/n_{s}=1+n_{s}\sum_{\bm{r}\neq0}\cos\left(\bm{k}\cdot\bm{r}\right)\left\{ \frac{1}{2\beta J(\bm{r})}-\coth\left[2\beta J(\bm{r})\right]\right\} .\label{eq:Sq_lowNS}
\end{equation}

For the parabolic dispersion, we can plug in the exact expression
(\ref{eq:rkky_Meijer}) into Eq.~(\ref{eq:Sq_lowNS}). Due to the $r^{-2}$
asymptotic decay of the RKKY interaction $J(\bm{r})$, the infinite
sum in (\ref{eq:Sq_lowNS}) converges rapidly as we increase the cutoff
$r<\Lambda$. For $\mu=-3.396t$, we obtain good convergence for $\Lambda\gtrsim256 a$
where $a$ is the lattice constant.

In Fig.~\ref{fig:Sq_lowNS}, we plot the static spin structure factor
obtained from Eq.~(\ref{eq:Sq_lowNS}) for the parabolic dispersion \eqref{eq:disp_parabolic}
with a low spin concentration  $n_s=0.2$. The result agrees qualitatively 
with the high-$T$ expansion at small $n_{s}$ [see Fig.~\ref{fig:Sq_highT}(g)], showing a clear enhancement
at $k=2k_{F}$ upon lowering temperature. This enhancement explains why the  
the introduction of a significant concentration of spin vacancies makes the resistivity upturn more pronounced.
As it is clear from Eq.~\eqref{tsc},  in the dilute limit the  two-spin correlator inherits the oscillatory nature of the real-space RKKY interaction. Consequently, its Fourier transform, given in Eq.~\eqref{eq:Sq_lowNS}, is strongly peaked at 
$2 k_F$.

Since Eq.~\eqref{eq:Sq_lowNS} only includes the  first two terms of an expansion in powers of $n_s$, this equation  is valid as long as the second term of order $\mathcal{O}(n_s)$ is much smaller than the first term. When this condition is violated at very low temperature $T \leq T_\text{low}$, the structure factor $\overline{\mathcal{S}}(\bm{k})$ is no longer positive-definite. For the typical $n_s$ values considered in this section ($0.01 \lesssim n_s \lesssim 0.2$), the lowest temperature for maintaining a positive-definite $\overline{\mathcal{S}}(\bm{k})$ is about $T_\text{low}/n_s \sim 0.005 J^2/t$.

For the parabolic dispersion, we obtain the following expression for the resistivity in the dilute limit:
\begin{equation}
\begin{split}
\frac{\rho}{n_{s}\rho_{0}} & =1+2n_{s}\sum_{\bm{r}\neq0}\left[\frac{J_{1}(2k_{F}r_{x})}{2k_{F}r_{x}}-J_{2}(2k_{F}r_{x})\right]\\
 & \quad\qquad\qquad\cdot\left\{ \frac{1}{2\beta J(\bm{r})}-\coth\left[2\beta J(\bm{r})\right]\right\} ,
\end{split}
\label{eq:rho_lowNS}
\end{equation}
where $J_{n}(x)$ are the Bessel functions of the first kind. Once again,
the infinite sum shows  good convergence for a cutoff $\Lambda\gtrsim256 a$.
The $\rho(T)$ curves shown in Fig.~\ref{fig:rho_lowNS} for different low-density values confirm the resistivity upturn. 
As  expected, the low-density approximation becomes more accurate as we decrease $n_s$ and increase $T$ (see comparison to the SLL simulation in Fig.~\ref{fig:rho_lowNS}).

The upturn remains noticeable only when the ratio between the magnetic impurity and the carrier concentration, $n_s/n_e$, is larger than a certain value.
This becomes clear if we rewrite Eq.~\eqref{eq:rho_lowNS} in the limit of $n_e \ll 1$ and $n_s \ll 1$:
\begin{equation}
\frac{\rho}{n_{s}\rho_{0}} \approx 1+ \frac{n_s}{n_e} g(\frac{tT }{J^2 n_e}),
\end{equation}
where $n_e\approx k_F^2/(2\pi)$ and
\begin{equation}
\begin{split}
g(y) &=\int_0^\infty r d r \left[ J_0^2(r)-J_1^2(r) \right] \\
&\quad \qquad \cdot \left\{ \frac{y}{2 {\tilde J}(r)}-\coth\left[\frac{2 {\tilde J}(r)}{y}\right]\right\}.
\end{split}
\end{equation}
We note that 
\begin{equation}
{\tilde J}(r) \equiv \frac{ 2 \pi t} {J^2 k_F^2}  J\left(\frac{r}{k_F}\right)
\end{equation}
is independent of $k_F$ according to Eq.~\eqref{eq:rkky_Meijer}. Given that
\begin{equation}
y= \frac{tT }{J^2 n_e} = \frac{tT}{J^2n_s} \frac{n_s}{n_e},
\end{equation}
it is clear that $y \ll 1$ for the base temperature $T_\text{low}/n_s\approx 0.005 J^2/t$ used in Fig.~\ref{fig:rho_lowNS}.
Given that $g(y) \approx 0.18/ \sqrt{y}$ for $y \ll 1$, we finally obtain
\begin{equation}
\frac{\rho}{n_{s}\rho_{0}} \approx 1+ 0.18 \sqrt{\frac{n_s}{n_e}} \sqrt{\frac{J^2 n_s}{tT }}.
\end{equation}
This equation implies that, given a dimensionless base temperature $\theta_B=T_\text{low}/n_s(J^2/t)$, the resistivity upturn becomes relatively small for 
$n_s/n_e \ll \theta_B$.


As already noted in Ref.~\citep{Rushbrooke1964}, the low-density
expansion can be obtained by rearranging terms in the high-$T$ expansion.
A straightforward expansion of Eq.~(\ref{eq:Sq_lowNS}) up to $\mathcal{O}(\beta^{3})$
reveals that
\begin{equation}
\overline{\mathcal{S}}(\bm{k})/n_{s}=1+K\tilde{\chi}_{\bm{k}}-\frac{3K^{3}}{5n_{s}^{2}}\frac{1}{N^{2}}\sum_{\bm{q}\bm{q}^{\prime}}\tilde{\chi}_{\bm{q}}\tilde{\chi}_{\bm{q}^{\prime}}\tilde{\chi}_{\bm{k}-\bm{q}-\bm{q}^{\prime}},
\end{equation}
where $K\equiv\frac{2\beta}{3}J^{2}n_{s}$. This expression coincides with 
the terms in Eq.~(\ref{eq:high-T}) up to $\mathcal{O}(n_{s}^{2})$.

\section{Effect of Magnetic Field\label{sec:magneto}}

As we discussed in the previous sections, the resistivity upturn considered
in this paper arises from the enhanced magnetic structure factor below
$2k_{F}$ as the temperature is lowered. However, it is clear that the ${\bm k}={\bm 0}$ component
of the structure factor does not contribute to the collision integral in Eq.~(\ref{eq:rho_born}).
Given that the application of a uniform magnetic field transfers spectral weight from finite ${\bm k}$ to
${\bm k}={\bm 0}$, the resulting magnetoresistance should be negative.

\begin{figure}
\centering
\includegraphics[width=1\columnwidth]{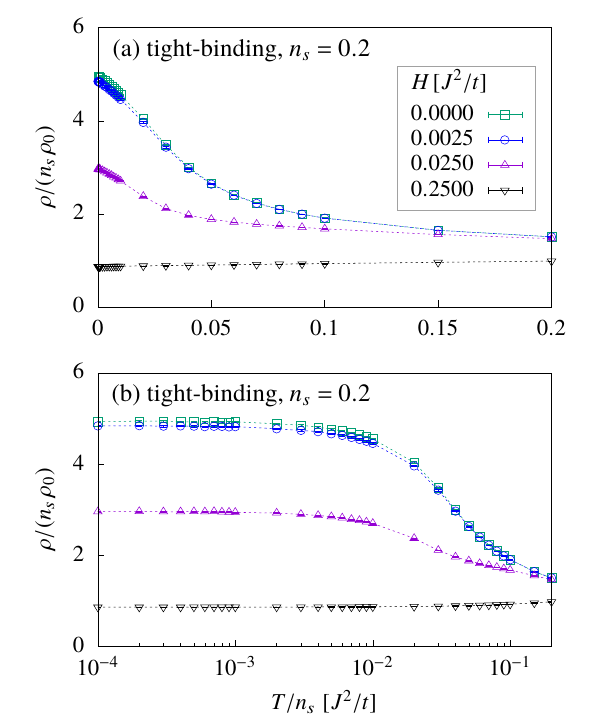}
\caption{Electrical resistivity obtained from SLL simulation of the RKKY model
on $L=128$ square lattice, for the susceptibility $\chi_{\bm{k}}^0$ obtained from the tight-binding dispersion 
\eqref{eq:disp_TB} and
different Zeeman fields $H=\{0,\,0.0025,\,0.025,\,0.25\}$
(unit: $J^{2}/t$). Both panels show the same data, with (a) linear
scale and (b) linear-logarithmic scale.\label{fig:Zeeman}}
\end{figure}

The Zeeman interaction between  uniform magnetic field and the magnetic impurities is:
\begin{equation}
\mathcal{H}_{\text{Zeeman}}=-H\sum_{i}\zeta_{i}S_{i}^{z},
\end{equation}
where we have absorbed the $g$-factor into the definition of $H$. The Zeeman coupling to itinerant electrons is not included because the Pauli susceptibility is several orders of magnitude 
smaller than the susceptibility of the local moments.

At low temperatures, the local moments become fully polarized  above the saturation field $H_{\rm sat}$ and
the electron-spin scattering is completely suppressed because
the spin structure factor has
a peak at $\bm{k}=0$ and negligible weight for  finite $\bm{k}$ values.
From Eq.~(\ref{eq:rho_born}), it is clear that such structure factor  cannot
produce a resistivity upturn upon lowering temperature.

For $H <  H_{\rm sat}$, a finite amount of the spectral 
weight in $\overline{\mathcal{S}}(\bm{k})$ is transferred from finite
$\bm{k}$ to $\bm{k}=0$. As shown in Fig.~\ref{fig:Zeeman}, this effect reduces the electron-spin
scattering along with the resistivity upturn. We note, however, that there  are other mechanisms, such as Kondo screening~\citep{Hewson97} and Anderson localization~\citep{Lee1985_rmp}, which also produce a negative magnetoresistance.

The above analysis, which leads to a negative magnetoresistance, is only appropriate for systems with isotropic magnetic interactions. The problem becomes more complex for materials with significant exchange or single-ion  anisotropy because the competition between the Zeeman and the anisotropy terms may induce nonlinear effects that  can  change the sign of magnetoresistance for magnetic fields well below saturation.
Nevertheless, regardless of the sign of the magnetoresistance, the RKKY-induced resistivity upturn should still survive, as long as $\overline{\mathcal{S}}(k \lesssim 2k_F)$ is enhanced upon lowering temperature.
We also note that the application of a magnetic field can also lead to more subtle effects, such as a nonmonotonic temperature dependence of the resistivity in RKKY systems with competing $k=2k_F$ and $k=0$ fluctuations, such as the one obtained for the tight-binding dispersion~\eqref{eq:disp_TB}, due to a  spectral weight redistribution  within the interval $0 \leq k \leq 2k_F$.

\section{Three-Dimensional Case \label{sec:ThreeD}}

In this section, we briefly extend our discussion to the 3D case.
For simplicity, we will  focus on the limit of low magnetic impurity concentration $n_{s}\ll1$,
and a low electron filling fraction  given by $k_F=0.777$. Consequently, as long as the band structure 
of the 3D lattice has a single global minimum (single electron pocket), we can approximate the single-electron dispersion with the  parabolic function
\begin{equation}
\epsilon_{\bm{k}}=\frac{k^{2}}{2m}.
\label{eq:disp_parabolic_3D}
\end{equation}
As  we will see below, the 3D case is much less sensitive to quartic or higher-order corrections because the 3D RKKY interaction is much less frustrated than the 2D case. 

\begin{figure}
\centering
\includegraphics[width=1\columnwidth]{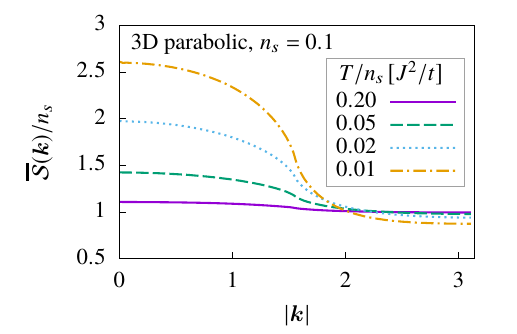}
\caption{Disorder averaged static spin structure factors obtained from the low-density
expansion (\ref{eq:Sq_lowNS}) for the 3D parabolic dispersion  \eqref{eq:disp_parabolic_3D} with
spin concentration $n_{s}=0.1$ and four different temperatures $T/n_{s}=\{0.2,0.05,0.02,0.01\}$
(unit: $J^{2}/t$). We set the Fermi wave vector to $k_{F}=0.777$.\label{fig:Sk_lowNS_3D}}

\centering
\includegraphics[width=1\columnwidth]{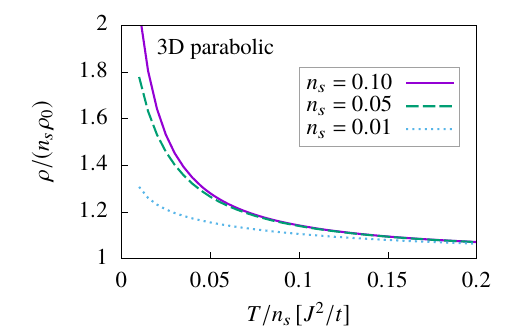}
\caption{Electrical resistivity obtained from the low-density expansion (\ref{eq:rho_lowNS_3d})
for the 3D parabolic dispersion \eqref{eq:disp_parabolic_3D} and three different spin concentrations
$n_{s}=\{0.1,0.05,0.01\}$. We set the Fermi wave vector $k_{F}=0.777$.\label{fig:rho_lowNS_3D}}
\end{figure}

The corresponding bare electronic susceptibility is again known analytically
at $T=0$:
\begin{equation}
\chi_{\bm{k}}^{0}=\frac{mk_{F}}{2\pi^{2}}\left[1+\frac{1-x^{2}}{2x}\ln\left|\frac{1+x}{1-x}\right|\right],\label{eq:chi_3D}
\end{equation}
where $x\equiv k/(2k_{F})$.

The RKKY interaction in real space is obtained by Fourier transforming \eqref{eq:chi_3D}:
\begin{equation}
J(\bm{r})=-\frac{J^{2}}{t}\,\frac{\sin(2k_{F}r)-2k_{F}r\cos(2k_{F}r)}{16\pi^{3}r^{4}}.\label{eq:rkky_real_3D}
\end{equation}
A big difference relative to the 2D case  is that the RKKY interaction is no longer highly frustrated
because $\chi_{\bm{k}}^{0}$ in Eq.~\eqref{eq:chi_3D} has a single global maximum at $\bm{k}=\bm{0}$ and it
decreases monotonically with $k$ for low electron filling. 

In the dense limit $n_s=1$, the low-temperature dependence of the resistivity should be affected by the onset of dominant ferromagnetic correlations. The situation is analogous to the 2D case with parabolic dispersion, where ferromagnetic correlations arise from an order by disorder mechanism. As we have seen in Sec.~\ref{sec:dense}, the onset of low-temperature ferromagnetic correlations at $T \sim T_F$ produces a maximum in the resistivity (see Fig.~\ref{fig:rho_SLL_parabolic}). In other words, the resistivity upturn stops around $T_F$. This is also the expected behavior of the resistivity for the 3D case
in the dense limit. The two main differences relative to the 2D case are the following. (i) The low-temperature ferromagnetic correlations of the 3D system  are more robust against small corrections of $\chi_{\bm{k}}^0$ due to lattice effects or the inclusion of electron-electron interactions. (ii) The onset of ferromagnetic correlations can in principle occur at a rather high temperature, making the resistivity upturn practically unnoticeable.
Both differences are a direct consequence of the much smaller frustration of the 3D RKKY interaction ($\chi_{\bm{k}}^{0}$ has a single global maximum).

What is the effect of dilution in the 3D case?
After replacing the 3D version of the RKKY interaction \eqref{eq:rkky_real_3D}  into the low-density expansion of $\overline{\mathcal{S}}(\bm{k})$ given in Eq.~(\ref{eq:Sq_lowNS}), it becomes clear that $\overline{\mathcal{S}}(\bm{k})$
still has a maximum at ${\bm k}={\bm 0}$ (see Fig.~\ref{fig:Sk_lowNS_3D}). The key observation, however, is that 
$\overline{\mathcal{S}}(\bm{k})$ remains close to $\overline{\mathcal{S}}(\bm{0})$ for $k\leq 2k_F$, while it drops to much smaller values for $k > 2k_F$. This behavior of $\overline{\mathcal{S}}(\bm{k})$ is enough to still produce a pronounced resistivity upturn because the higher dimensionality enlarges the relative volume of the phase space that has a significant contribution to the collision integral. In other words, the weight factor, $x^2/\sqrt{1-x^2}$ of the 2D collision integral \eqref{eq:rho_born} is replaced by $x^3$ in the 3D collision integral:
\begin{equation}
\rho=4\rho_{0}\int_{0}^{1} d x\overline{\mathcal{S}}(2k_{F}x)x^{3},\label{eq:rho_born_3D}
\end{equation}
where $\rho_{0}=3\pi J^{2}/(tek_{F})^{2}$. In 2D, the dominant contribution to the resistivity comes mostly from back scattering processes ($k$ very close to $2k_F$) because they are the ones that produce the maximal deviation angle ($\theta=\pi$) from the original direction of propagation of the conduction electron (the differential scattering cross section $\sigma(\theta)$ is multiplied by a factor $(1-\cos{\theta})$). In 3D, the relative weight of the scattering processes decreases more slowly as we move away from the back scattering condition ($\theta=\pi$ or $k=2k_F$) because the number of final states, which is proportional to $\sin{\theta}$ increases  and reaches its maximum value for $\theta=\pi/2$ or $k=\sqrt{2} k_F$  (the differential scattering cross section $\sigma(\theta)$ is multiplied by a factor $(1-\cos{\theta}) \sin{\theta}$).
Consequently, the temperature dependence of $\overline{\mathcal{S}}(\bm{k})$ shown in Fig.~\ref{fig:Sk_lowNS_3D} is enough to still produce a pronounced resistivity upturn.

Once again, we emphasize that  the enhancement of $\overline{\mathcal{S}}(\bm{k})$ below $2k_{F}$ in the dilute limit under consideration ($n_{s} \ll 1$) is a direct consequence of the momentum dependence of the RKKY interaction given in Eq.~\eqref{eq:chi_3D}. After inserting  the 3D expression of  $\overline{\mathcal{S}}(\bm{k})$ that is obtained from Eqs.~\eqref{eq:rkky_real_3D} and \eqref{eq:Sq_lowNS} into the 3D collision integral \eqref{eq:rho_born_3D}, we obtain
\begin{widetext}
\begin{equation}
\frac{\rho}{n_{s}\rho_{0}}=1+4n_{s}\sum_{\bm{r}\neq0}\frac{6+3\left[\left(2k_{F}r_{x}\right)^{2}-2\right]\cos\left(2k_{F}r_{x}\right)+2k_{F}r_{x}\left[\left(2k_{F}r_{x}\right)^{2}-6\right]\sin\left(2k_{F}r_{x}\right)}{\left(2k_{F}r_{x}\right)^{4}}\left\{ \frac{1}{2\beta J(\bm{r})}-\coth\left[2\beta J(\bm{r})\right]\right\} .\label{eq:rho_lowNS_3d}
\end{equation}
\end{widetext}

Figure~\ref{fig:rho_lowNS_3D} shows the resistivity curves obtained for three different values of the density $n_{s}=\{0.1,0.05,0.01\}$. As expected, there is a clear resistivity upturn that, like in the 2D case, becomes less pronounced 
for small values of $n_s/n_e$.


\section{Conclusions \label{sec:conclusion}}

To summarize, the resistivity upturn that was reported in the dense limit $n_s=1$
for a highly frustrated  RKKY interaction~\citep{Wang2016_resistivity} becomes even more pronounced
for $n_e < n_s < 1$. The disorder introduced by the dilution effect eliminates the requirement of magnetic frustration for the stabilization of classical spin liquid with dominant antiferromagnetic correlations at $k \lesssim 2k_F$. Moreover, unlike the dense limit case,  this classical spin liquid regime induced by disorder persists down to $T=0$. Since magnetic frustration is no longer required, the resistivity upturn is also found in the 3D case, where the RKKY interaction favors ferromagnetic correlations.
In other words, by moving away from the dense limit, the resistivity upturn induced by the RKKY interaction becomes a more  general effect.

It is important to keep in mind that the dilution effect reduces the energy scale of the RKKY interaction: $T_\text{RKKY} \to n_s T_\text{RKKY}$. It is then clear that for antiferromagnetic interaction between the local moments and the conduction electrons, the Kondo temperature will become larger than $n_s T_\text{RKKY}$ for low enough values of $n_s$. As we have shown in this manuscript, another limiting factor is the carrier concentration $n_e$: the resistivity upturn induced by the RKKY interaction remains noticeable only for $n_s/n_e$ larger than a certain value. For 3D systems, it has been suggested that the combination of disorder, frustration, and anisotropy in the dilute RKKY systems leads to a glass transition at a low enough temperature $T_G$~\cite{Fischer_book}. Such transition has not been considered here because a numerical study of dilute  3D systems, including possible effects of magnetic anisotropy, is beyond the scope of this manuscript. We conjecture that the resistivity upturn that is reported here for the 3D case should saturate for $T<T_G$ and exhibit hysteric behavior
in field-cooled and zero-field-cooled experiments or simulations.

Finally, for the dense limit ($n_s=1$), the highly frustrated nature of the RKKY interaction generated by a 2D electron gas makes the low-temperature transport properties  very sensitive to small lattice effects. 
For the parabolic dispersion of the 2D electron gas, the resistivity upturn is finally suppressed at low enough temperatures
due to a ferromagnetic tendency  induced by an order by disorder mechanism. In contrast, lattice effects (tight-binding dispersion for nearest-neighbor hopping) induce a negative quartic correction to the single-electron dispersion, which
leads to  bond-density wave ordering (translation and $C_{4}$ rotation symmetry breaking). The bond-density wave ordering does not suppress the resistivity upturn. We note that higher-order corrections to the parabolic dispersion
are not the only factors that can lift the degeneracy of  susceptibility $\chi_{\bm{k}}^0$ of the 2D electron gas.
The inclusion of  electron-electron interactions  also modify the momentum dependence of  $\chi_{\bm{k}}^0$, 
although it is not fully settled what is the magnitude of the wave vector that maximizes the susceptibility~\citep{Simon2008}. Other types
of exchange interactions between the local moments, including direct exchange,
super-exchange and dipole-dipole interactions, can also lift the degeneracy of $\chi_{\bm{k}}^0$ for $k\leq 2 k_F$ and modify the low-temperature behavior of the resistivity curve. Consequently, the results presented in this paper for the RKKY interaction derived from the tight-binding dispersion \eqref{eq:disp_TB} are representative of a more general physical situation represented by the magnetic susceptibility $\chi_{\bm{k}}^0$ shown in Fig.~\ref{fig:chi0}~(b).

\begin{acknowledgments}
We thank K.~Barros, D.~Maslov, F.~Ronning, and H.~Suwa for helpful discussions.
Z.\nobreak\,W. and C.\nobreak\,D.\nobreak\,B. are supported by funding from the Lincoln Chair of Excellence in Physics.
This research used resources
of the National Energy Research Scientific Computing Center (NERSC),
a U.S. Department of Energy Office of Science User Facility operated
under Contract No. DE-AC02-05CH11231. 
This research used resources
of the Oak Ridge Leadership Computing Facility, which is a DOE Office
of Science User Facility supported under Contract DE-AC05-00OR22725.
\end{acknowledgments}

\appendix

\begin{figure}
	\centering
	\includegraphics[width=\columnwidth]{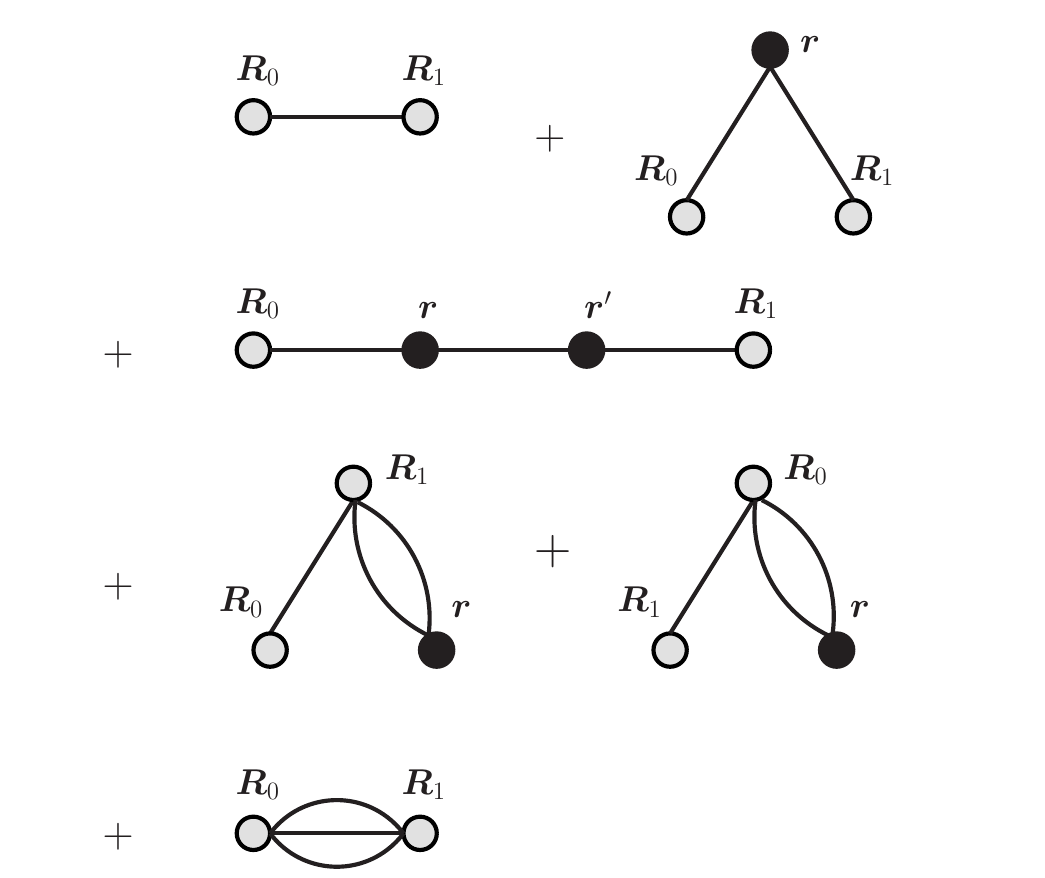}
	\caption{Nonzero free graphs of spin correlator $\langle\bm{S}_{\bm{R}_{0}}\cdot\bm{S}_{\bm{R}_{1}}\rangle$
		for the RKKY model (\ref{eq:RKKY}) under zero magnetic field, up
		to $\mathcal{O}(\beta^{3})$.\label{fig:free_graph}}
\end{figure}

\section{High-Temperature Expansion\label{sec:High-T}}
The two-spin correlator at relatively high temperatures can be obtained
from the standard free graph expansion~\citep{Oitmaa_book}. In absence 
of external magnetic field, the only nonzero graphs for the RKKY
Hamiltonian (\ref{eq:RKKY}) up to $\mathcal{O}(\beta^{3})$ are given
by Fig.~\ref{fig:free_graph}. For $\bm{R}_{0}\neq\bm{R}_{1}$ on
a square lattice:\begin{widetext}
\begin{align}
\langle\bm{S}_{\bm{R}_{0}}\cdot\bm{S}_{\bm{R}_{1}}\rangle
 & =-\frac{2\beta}{3}\zeta_{\bm{R}_{0}}\zeta_{\bm{R}_{1}}J(\bm{R}_{0}-\bm{R}_{1}) +\frac{4\beta^{2}}{9}\zeta_{\bm{R}_{0}}\zeta_{\bm{R}_{1}}\sum_{\bm{r}}\zeta_{\bm{r}}J(\bm{R}_{0}-\bm{r})J(\bm{R}_{1}-\bm{r}) \nonumber \\
 & \quad-\frac{8\beta^{3}}{27}\zeta_{\bm{R}_{0}}\zeta_{\bm{R}_{1}}\sum_{\bm{r}\bm{r}^{\prime}}\zeta_{\bm{r}}\zeta_{\bm{r}^{\prime}}J(\bm{R}_{0}-\bm{r})J(\bm{R}_{1}-\bm{r}^{\prime})J(\bm{r}-\bm{r}^{\prime})\nonumber \\
 & \quad+\frac{8\beta^{3}}{27}\zeta_{\bm{R}_{0}}\zeta_{\bm{R}_{1}}J(\bm{R}_{0}-\bm{R}_{1})\sum_{\bm{r}}\zeta_{\bm{r}}\left[J^{2}(\bm{R}_{0}-\bm{r})+J^{2}(\bm{R}_{1}-\bm{r})\right]\nonumber \\
 & \quad-\frac{16\beta^{3}}{135}\zeta_{\bm{R}_{0}}\zeta_{\bm{R}_{1}}J^{3}(\bm{R}_{0}-\bm{R}_{1})+\mathcal{O}(\beta^{4}).
\end{align}

By taking the Fourier transform (\ref{eq:Fourier_S}) and using
\begin{equation}
\tilde{\delta}_{\bm{k}}\equiv\frac{1}{N}\sum_{\bm{r}}\zeta_{\bm{r}}e^{\iu\bm{k}\cdot\bm{r}},
\end{equation}
we obtain the static spin structure factor in $\bm{k}$-space:
\begin{align}
\mathcal{S}(\bm{k}) & =\tilde{\delta}_{\bm{0}}+\frac{2\beta}{3}J^{2}\sum_{\bm{q}}\left|\tilde{\delta}_{\bm{k}+\bm{q}}\right|^{2}\tilde{\chi}_{\bm{q}}+\frac{4\beta^{2}}{9}J^{4}\sum_{\bm{q}\bm{q}^{\prime}}\Big(\tilde{\delta}_{\bm{q}}\tilde{\delta}_{\bm{q}^{\prime}}\tilde{\delta}_{-\bm{q}-\bm{q}^{\prime}}\tilde{\chi}_{\bm{k}-\bm{q}}\tilde{\chi}_{\bm{k}+\bm{q}^{\prime}}-\frac{1}{N}\left|\tilde{\delta}_{\bm{q}}\right|^{2}\tilde{\chi}_{\bm{q}^{\prime}}\tilde{\chi}_{\bm{q}-\bm{q}^{\prime}}\Big)\nonumber \\
 & \quad+\frac{8\beta^{3}}{27}J^{6}\sum_{\bm{k}_{1}\bm{k}_{2}\bm{k}_{3}}\left[\tilde{\delta}_{\bm{k}+\bm{k}_{1}}\tilde{\delta}_{-\bm{k}+\bm{k}_{2}}\tilde{\delta}_{-\bm{k}_{1}+\bm{k}_{3}}\tilde{\delta}_{-\bm{k}_{2}-\bm{k}_{3}}\tilde{\chi}_{\bm{k}_{1}}\tilde{\chi}_{\bm{k}_{2}}\tilde{\chi}_{\bm{k}_{3}}-\frac{1}{N}\tilde{\delta}_{\bm{k}_{1}+\bm{k}_{2}}\tilde{\delta}_{-\bm{k}_{1}+\bm{k}_{3}}\tilde{\delta}_{-\bm{k}_{2}-\bm{k}_{3}}\tilde{\chi}_{\bm{k}_{1}}\tilde{\chi}_{\bm{k}_{2}}\tilde{\chi}_{\bm{k}_{3}}\right]\nonumber \\
 & \quad-\frac{8\beta^{3}}{27}J^{6}\frac{1}{N}\sum_{\bm{k}_{1}\bm{k}_{2}\bm{k}_{3}}\left(\tilde{\delta}_{\bm{k}+\bm{k}_{1}+\bm{k}_{2}+\bm{k}_{3}}\tilde{\delta}_{-\bm{k}-\bm{k}_{1}}\tilde{\delta}_{-\bm{k}_{2}-\bm{k}_{3}}+\tilde{\delta}_{\bm{k}+\bm{k}_{1}}\tilde{\delta}_{-\bm{k}-\bm{k}_{1}+\bm{k}_{2}+\bm{k}_{3}}\tilde{\delta}_{-\bm{k}_{2}-\bm{k}_{3}}\right)\tilde{\chi}_{\bm{k}_{1}}\tilde{\chi}_{\bm{k}_{2}}\tilde{\chi}_{\bm{k}_{3}}\nonumber \\
 & \quad+\frac{16\beta^{3}}{135}J^{6}\frac{1}{N^{2}}\sum_{\bm{k}_{1}\bm{k}_{2}\bm{k}_{3}}\left|\tilde{\delta}_{\bm{k}+\bm{k}_{1}+\bm{k}_{2}+\bm{k}_{3}}\right|^{2}\tilde{\chi}_{\bm{k}_{1}}\tilde{\chi}_{\bm{k}_{2}}\tilde{\chi}_{\bm{k}_{3}}+\mathcal{O}(\beta^{4}).
\end{align}

These are some useful formulas for the disorder average:
\begin{subequations}
\begin{align}
\overline{\left|\tilde{\delta}_{\bm{k}}\right|^{2}} & =n_{s}^{2}\delta_{\bm{k}}+\frac{n_{s}(1-n_{s})}{N},\\
\overline{\tilde{\delta}_{\bm{q}}\tilde{\delta}_{\bm{q}^{\prime}}\tilde{\delta}_{-\bm{q}-\bm{q}^{\prime}}} & =n_{s}^{3}\delta_{\bm{q}}\delta_{\bm{q}^{\prime}}+\frac{n_{s}^{2}(1-n_{s})}{N}\left(\delta_{\bm{q}+\bm{q}^{\prime}}+\delta_{\bm{q}^{\prime}}+\delta_{\bm{q}}\right)+\frac{n_{s}(1-n_{s})(1-2n_{s})}{N^{2}},\\
\overline{\tilde{\delta}_{\bm{k}+\bm{k}_{1}}\tilde{\delta}_{-\bm{k}+\bm{k}_{2}}\tilde{\delta}_{-\bm{k}_{1}+\bm{k}_{3}}\tilde{\delta}_{-\bm{k}_{2}-\bm{k}_{3}}} & =n_{s}^{4}\delta_{\bm{k}+\bm{k}_{1}}\delta_{-\bm{k}+\bm{k}_{2}}\delta_{-\bm{k}_{1}+\bm{k}_{3}}\nonumber \\
 & \quad+\frac{n_{s}^{3}(1-n_{s})}{N}\Big(\delta_{\bm{k}+\bm{k}_{1}}\delta_{-\bm{k}+\bm{k}_{2}}+\delta_{\bm{k}+\bm{k}_{1}}\delta_{-\bm{k}-\bm{k}_{3}}+\delta_{\bm{k}+\bm{k}_{1}}\delta_{-\bm{k}_{2}-\bm{k}_{3}}\nonumber \\
 & \quad\qquad\qquad\quad+\delta_{-\bm{k}+\bm{k}_{2}}\delta_{-\bm{k}_{1}+\bm{k}_{3}}+\delta_{\bm{k}+\bm{k}_{3}}\delta_{-\bm{k}+\bm{k}_{2}}+\delta_{\bm{k}_{1}+\bm{k}_{2}}\delta_{-\bm{k}_{1}+\bm{k}_{3}}\Big)\nonumber \\
 & \quad+\frac{n_{s}^{2}(1-n_{s})(1-2n_{s})}{N^{2}}\left(\delta_{\bm{k}+\bm{k}_{1}}+\delta_{\bm{k}-\bm{k}_{2}}+\delta_{\bm{k}_{1}-\bm{k}_{3}}+\delta_{\bm{k}_{2}+\bm{k}_{3}}\right)\nonumber \\
 & \quad+\frac{n_{s}^{2}(1-n_{s})^{2}}{N^{2}}\left(\delta_{\bm{k}_{1}+\bm{k}_{2}}+\delta_{\bm{k}+\bm{k}_{3}}+\delta_{\bm{k}+\bm{k}_{1}-\bm{k}_{2}-\bm{k}_{3}}\right)+\frac{n_{s}}{N^{3}}(1-n_{s})(1-6n_{s}+6n_{s}^{2}).
\end{align}
\end{subequations}

The disorder averaged static spin structure factor is
\begin{align}
\overline{\mathcal{S}}(\bm{k})/n_{s} & =1+K\tilde{\chi}_{\bm{k}}+K^{2}\left(\tilde{\chi}_{\bm{k}}^{2}-\frac{1}{N}\sum_{\bm{q}}\tilde{\chi}_{\bm{q}}^{2}\right)\nonumber \\
 & \quad+K^{3}\left[\tilde{\chi}_{\bm{k}}^{3}-\frac{1}{N}\sum_{\bm{q}}\tilde{\chi}_{\bm{q}}^{3}-\frac{2}{N}\tilde{\chi}_{\bm{k}}\sum_{\bm{q}}\tilde{\chi}_{\bm{q}}^{2}+\left(1-\frac{3}{5n_{s}^{2}}\right)\frac{1}{N^{2}}\sum_{\bm{q}\bm{q}^{\prime}}\tilde{\chi}_{\bm{q}}\tilde{\chi}_{\bm{q}^{\prime}}\tilde{\chi}_{\bm{k}-\bm{q}-\bm{q}^{\prime}}\right]+\mathcal{O}(\beta^4),
\end{align}
where $K\equiv\frac{2\beta}{3}J^{2}n_{s}$.
\end{widetext}

\section{Stochastic Landau-Lifshitz Dynamics\label{sec:SLL}}
The classical (large-$S$) limit of the RKKY model~(\ref{eq:RKKY}) can be studied
by unbiased numerical methods. Typical choices include classical Monte
Carlo (MC) and the SLL dynamics, which give the same exact results on finite lattices
for models without large spin anisotropy. Here we use SLL dynamics because it is numerically more efficient.

The SLL equation is
\begin{equation}
\frac{d\bm{S}_{i}}{d\tau}=-\gamma\bm{S}_{i}\times\left(\bm{f}_{i}+\bm{b}_{i}\right)-\frac{\alpha\gamma}{S}\bm{S}_{i}\times\left[\bm{S}_{i}\times\left(\bm{f}_{i}+\bm{b}_{i}\right)\right],\label{eq:SLL}
\end{equation}
where $\bm{f}_{i}$ are the molecular fields produced by the RKKY interactions:
\begin{equation}
\bm{f}_{i}=-\frac{d H_{\text{RKKY}}}{d\bm{S}_{i}}=2\frac{J^{2}}{\sqrt{N}}\sum_{\bm{k}}e^{\iu\bm{k}\cdot\bm{r}_{i}}\tilde{\chi}_{\bm{k}}\bm{S}_{-\bm{k}},\label{eq:force}
\end{equation}
and $\bm{b}_{i}$ are Gaussian random fields which satisfy
\begin{subequations}
\begin{align}
\langle\bm{b}_{i}(\tau)\rangle & =0,\\
\langle b_{i,\alpha}(\tau)b_{j,\beta}(0)\rangle & =2D_{\text{LL}}\delta_{ij}\delta_{\alpha\beta}\delta(\tau).
\end{align}
\end{subequations}

The value of $D_{\text{LL}}$ is fixed by the fluctuation-dissipation
theorem,
\begin{equation}
D_{\text{LL}}=\frac{\alpha}{1+\alpha^{2}}\frac{\bm{k}_{B}T}{\gamma S}.
\end{equation}

The gyromagnetic ratio and the damping factor are set to $\gamma=1$ and
$\alpha=1$, respectively.
Note that the fields $\bm{f}_{i}$ can be obtained by two consecutive
Fast Fourier Transforms (FFT). Thus, the cost of computing $\bm{f}_{i}$
on all sites is $\mathcal{O}(N\ln N)$, which is an important gain in comparison to
naive $\mathcal{O}(N^{2})$ implementations.

Denote the r.h.s of Eq.~(\ref{eq:SLL}) as $\bm{F}_{i}(\tau,\{\bm{S}_{j}(\tau)\})$.
From a random initial spin configuration at time $\tau=0$, we numerically
solve Eq.~(\ref{eq:SLL}) by the Euler predictor-corrector (Heun)
method:\begin{subequations}
\begin{align}
\bar{\bm{S}}_{i}(\tau_{n+1}) & =\bm{S}_{i}(\tau_{n})+\delta\tau\bm{F}_{i}\left(\tau_{n},\left\{ \bm{S}_{j}(\tau_{n})\right\} \right),\\
\bm{S}_{i}(\tau_{n+1}) & =\bm{S}_{i}(\tau_{n})+\frac{\delta\tau}{2}\bm{F}_{i}\left(\tau_{n},\left\{ \bm{S}_{j}(\tau_{n})\right\} \right)\nonumber \\
 & \quad\qquad~~+\frac{\delta\tau}{2}\bm{F}_{i}\left(\tau_{n+1},\left\{ \bar{\bm{S}}_{j}(\tau_{n+1})\right\} \right),
\end{align}
\end{subequations}where $\bar{\bm{S}}_{i}(\tau_{n+1})$ is the predictor
and $\bm{S}_{i}(\tau_{n+1})$ is the corrected new spin configuration.
After each step, we renormalize the spin magnitude to $S=1$.

Throughout this paper, we use a conservative time step $n_{s}\delta\tau=0.1\left(J^{2}/t\right)^{-1}$
(this choice has been benchmarked against MC simulations). We use time $\tau_{\text{eq}}$
to equilibrate the system, and another
$8\tau_{\text{eq}}$ for measurements.
The equilibration times $\tau_{\text{eq}}$ used in the SLL simulation
for different system sizes are:
(1) $L=32$: $\tau_{\text{eq}} = 2\times10^{4} \left(J^{2}/t\right)^{-1}$;
(2) $L=64$: $\tau_{\text{eq}} = 1\times10^{5} \left(J^{2}/t\right)^{-1}$;
and (3) $L=128$: $\tau_{\text{eq}} = 4\times10^{5} \left(J^{2}/t\right)^{-1}$.

Since we only have discrete values of momenta on finite lattices, we
estimate the Born approximation by a Lorentzian broadening of the
delta functions. Equations (\ref{eq:relaxation_time}) and (\ref{eq:rho_born})
become
\begin{align}
\frac{1}{\tau_{\bm{k}_{F}}} & \approx\frac{4\pi J^{2}}{N}\sum_{\bm{k}}\frac{\eta}{\pi\left[\left(\mu-\epsilon_{\bm{k}_{F}+\bm{k}}\right)^{2}+\eta^{2}\right]}\overline{\mathcal{S}}(\bm{k})\nonumber \\
 & \quad\qquad\qquad\times\left(1-\cos\theta_{\bm{k}_{F},\bm{k}_{F}+\bm{k}}\right),\\
\rho & =\frac{4}{N}\rho_{0}\sum_{\bm{k}}\frac{\eta t}{\left(\epsilon_{\bm{k}_{F}}-\epsilon_{\bm{k}_{F}+\bm{k}}\right)^{2}+\eta^{2}}\overline{\mathcal{S}}(\bm{k})\nonumber \\
 & \quad\qquad\qquad\times\left(1-\cos\theta_{\bm{k}_{F},\bm{k}_{F}+\bm{k}}\right),
\end{align}
where the sum is performed for $\overline{\mathcal{S}}(\bm{k})$ for
every discrete $\bm{k}$ point in the 1st BZ. Throughout the paper,
we choose the broadening factor $\eta=t/L$.

In principle, any $\bm{k}_{F}$ on the small FS can be used for evaluating the
resistivity. In Figs.~\ref{fig:rho_SLL_parabolic} and
\ref{fig:rho_SLL_square}, we average
the resistivity over 360 different $\bm{k}_{F}$ uniformly distributed
on the FS to achieve better statistics. In Fig.~\ref{fig:nematic}, we use $\bm{k}_{F}=(k_{F},0)$
for evaluating $\rho_{xx}$ and $\bm{k}_{F}=(0,k_{F})$ for evaluating
$\rho_{yy}$.

\section{Bond-density wave phase\label{sec:bond-density}}
\begin{figure}
\centering
\includegraphics[width=1\columnwidth]{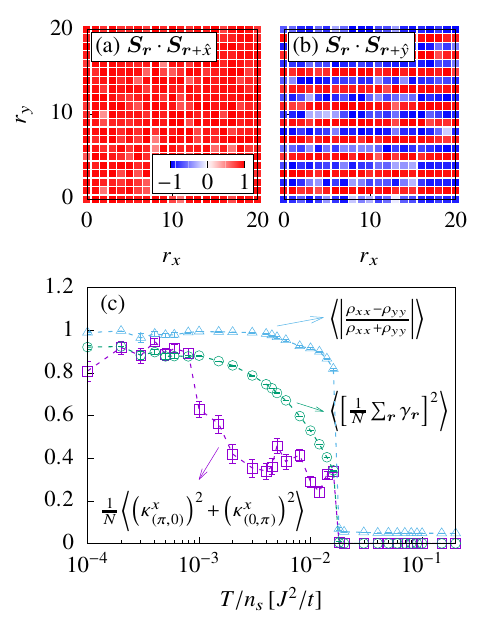}
\caption{SLL simulation results of the RKKY model on a square lattice
of $128 \times 128$ sites for $\chi^0_{\bm{k}}$ obtained from the tight-binding dispersion and spin concentration $n_{s}=1$. [(a) and (b)]
Snapshot of the local bond order parameters at temperature $T=0.005J^{2}/t$.
Note that these panels include only a small region of the finite square lattice that has been used in the simulation. 
(c) Temperature dependence of the bond order parameter, Ising nematic order parameter, and
resistivity anisotropy. The error bars
are estimated by averaging over 64 disorder realizations. \label{fig:nematic}}
\end{figure}

In this section, we will discuss the low-temperature ordered phase of the dense limit $n_{s}=1$. As it is shown in Fig.~\ref{fig:chi0}(c), $\chi_{\bm{k}}^{0}$ has a very small  angular dependence for the tight-binding dispersion on the square lattice, with four degenerate maxima at the discrete wave vectors: $\pm\bm{Q}_{1}\equiv\pm(2k_{F},0)$ and $\pm\bm{Q}_{2}\equiv\pm(0,2k_{F})$. At high temperatures, the magnetic structure factor has the same intensity at these  four wave vectors because they are related by symmetry transformations (diagonal reflections and 90$^\circ$ rotations). However, as it is shown  in the inset of Fig.~\ref{fig:Sq_SLL}(b), these symmetries are spontaneously broken at low enough temperatures: $\overline{\mathcal{S}}(\bm{k})$ has strong intensity either at 
$\pm\bm{Q}_{1}$ or at $\pm\bm{Q}_{2}$.\footnote{We have also verified this transition using a Monte Carlo simulation.}
The choice of $\pm\bm{Q}_{1}$ or $\pm\bm{Q}_{2}$ depends on the initial random spin configuration (the tunneling time between both configurations becomes much longer than the simulation time at low  enough temperatures).

The low-temperature phase breaks both the lattice $C_{4}$ symmetry
and the translational symmetry of the original Hamiltonian. To characterize
this phase, we introduce the local bond order parameters
\begin{subequations}
\begin{align}
\kappa_{\bm{r}}^{x} & \equiv\bm{S}_{\bm{r}}\cdot\bm{S}_{\bm{r}+\hat{x}},\\
\kappa_{\bm{r}}^{y} & \equiv\bm{S}_{\bm{r}}\cdot\bm{S}_{\bm{r}+\hat{y}},
\end{align}
\end{subequations} and the local Ising nematic order parameter,
\begin{equation}
\gamma_{\bm{r}}\equiv\bm{S}_{\bm{r}}\cdot\bm{S}_{\bm{r}+\hat{x}}-\bm{S}_{\bm{r}}\cdot\bm{S}_{\bm{r}+\hat{y}}.
\end{equation}

Figures \ref{fig:nematic}(a) and \ref{fig:nematic}(b) show the real-space distribution
of $\kappa_{\bm{r}}^{x}$ and $\kappa_{\bm{r}}^{y}$ for a snapshot
of the SLL simulation. Since the magnetic susceptibility is maximized at $2 k_F$,
the bond ordering wave number
is expected to be $4k_{F}$. For the finite size systems that have been
simulated, the bond ordering wave vector locks at  the closest value to $4k_{F}$ that
is commensurate with the lattice. For the particular case
$\mu=-3.396t$, we get $4k_{F}\approx3.193$. 
For finite square lattices of linear size up to $L=128$, the bond ordering wave vectors
that result from the SSL simulations are always $(\pi,0)$ or $(0,\pi)$.
For this reason, we use bond susceptibility
\begin{equation}
\frac{1}{N}\left\langle \left(\kappa_{(\pi,0)}^{x}\right)^{2}+\left(\kappa_{(0,\pi)}^{y}\right)^{2}\right\rangle 
\end{equation}
to identify the transition shown in Fig.~\ref{fig:nematic}, where $\kappa_{\bm{k}}^{x/y}$
are the Fourier transforms of $\kappa_{\bm{r}}^{x/y}$. The transition
temperature is found to be $T_{c}\approx0.002J^{2}/t$. The corresponding
error bars in Fig.~\ref{fig:nematic} are quite large because of the
the multiple domains that result from  the SLL simulations.

The $C_{4}$ rotation symmetry breaking
is revealed by the Ising-nematic order parameter
\begin{equation}
\left\langle \left[\frac{1}{N}\sum_{\bm{r}}\gamma_{\bm{r}}\right]^{2}\right\rangle 
\end{equation}
shown in Fig.~\ref{fig:nematic}(c) and by the resulting anisotropy of  the resistivity  
\begin{equation}
\left\langle \left|\frac{\rho_{xx}-\rho_{yy}}{\rho_{xx}+\rho_{yy}}\right|\right\rangle.
\end{equation}

The parabolic dispersion does not contain the $C_4$ lattice anisotropy.
Our SLL simulations for the magnetic susceptibility $\chi^0_{\bm{k}}$
obtained from the parabolic dispersion do not exhibit any nematic transition  down to $T/n_{s}=10^{-4}J^{2}/t$.
This result indicates that the lattice anisotropy plays a crucial role in the stabilization of the bond ordering.

\section{Finite Size Effects in SLL Simulations\label{sec:Finite-Size}}
\begin{figure}
	\centering
	\includegraphics[width=1\columnwidth]{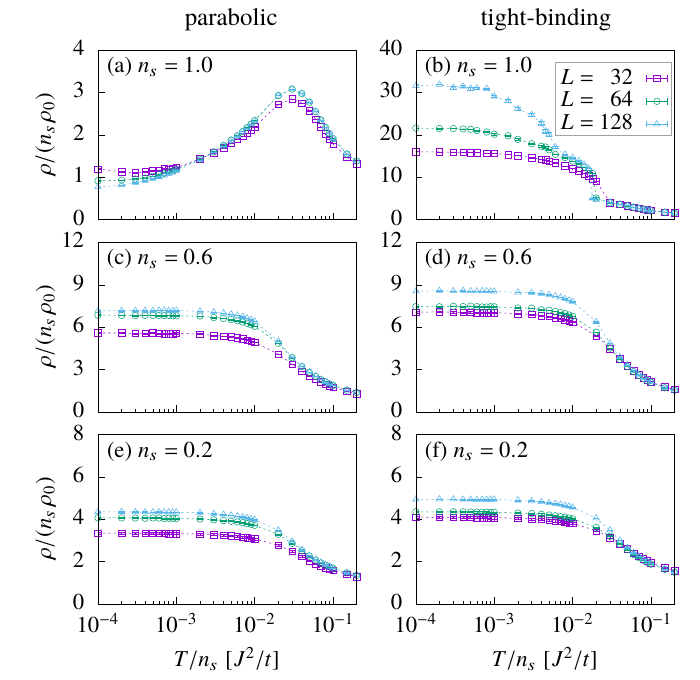}
	\caption{Electrical resistivity obtained from SLL simulation of the RKKY model
		on square lattice with different system sizes $L=\{32,\,64,\,128\}$,
		for parabolic dispersion (left column) and tight-binding dispersion
		(right column). The errorbars are estimated by averaging over 64 disorder
		realizations.\label{fig:finite_size}}
\end{figure}

In Fig.~\ref{fig:finite_size}, we compare the SLL results for three
different system sizes $L=\{32,64,128\}$. The results are qualitatively
the same for different system sizes. A relatively good convergence
is achieved in most of the cases, except for the dense limit ($n_{s}=1$) of the tight-binding 
case {[}see Fig.~\ref{fig:finite_size}(b){]}, which exhibits the
bond-density wave ordering at low temperatures.

The formation of the bond-density wave phase at low temperatures is accompanied by the generation of metastable domain wall defects
in our SLL simulation (we initialize
the spins from random configurations in all of our SLL simulations).
These defects are very difficult to eliminate within a reasonable
amount of computation time (unless we bias the initial spin
configuration as the bond-density wave with one domain, or use some
advanced MC update schemes). This is partly the reason of the slow
convergence of the tight-binding case for $n_{s}=1$ plot {[}see Fig.~\ref{fig:finite_size}(b){]}.

There is still another finite size effect that we noticed in our SLL simulation
for the tight-binding dispersion and $n_{s}=1$. 
The real-space molecular fields given in Eq.~(\ref{eq:force})
are obtained from a sum over $N=L^{2}$ wave vectors  $\bm{k}$. In the
thermodynamic limit, the peaks of $\tilde{\chi}_{\bm{k}}$ are located
exactly at $\pm(2k_{F},0)$ and $\pm(0,2k_{F})$ {[}see Fig.~\ref{fig:chi0}(c){]}.
However, these points cannot be accessed on a finite lattice without fine-tuning the chemical potential.
Thus, the maxima of $\tilde{\chi}_{\bm{k}}$ are shifted slightly
off the high-symmetry directions.
The SLL simulations pick up this finite size effect at the lowest temperatures by showing peaks slightly off the high-symmetry axes.

\bibliographystyle{apsrev4-1}
\bibliography{ref}

\end{document}